\documentclass[12pt]{article}
\usepackage[utf8]{inputenc}
\usepackage[T1]{fontenc}
\usepackage{mathtools}
\usepackage{ragged2e}
\usepackage{tabularx}
\usepackage{color,soul}
\usepackage{eurosym}
\usepackage[english]{babel}
\usepackage[margin=1in]{geometry}
\usepackage{amsmath,amsthm,amssymb}
\usepackage{indentfirst}
\usepackage{hyperref}
\usepackage{ulem}
\usepackage{setspace}
\usepackage{adjustbox}
\usepackage{rotating}       % for sidewaystable
\usepackage{threeparttable} % for nice table notes
\usepackage{subcaption}         % for subfloat
\usepackage{booktabs}       % for improved rules
\usepackage{array}
\usepackage{float} 
\usepackage[]{caption}
\usepackage[justification=justified,singlelinecheck=false, font={small},labelfont={bf}]{caption}
\usepackage{booktabs}    % For professional-looking rules
\usepackage{siunitx}     % For numeric alignment
\usepackage{subcaption}
\newcommand{\setfoot}[2]{%
    \footnote{#2}%
    \newcounter{#1}%
    \setcounter{#1}{\value{footnote}}%
}

\newcommand{\qedwhite}{\hfill \ensuremath{\Box}}

\newcommand\deltameq[1]%
  {\stackrel{\scriptscriptstyle(\mkern-1.5mu#1\mkern-1.5mu)}{=}}

\title{Linear Demand Systems are Consistent with Discrete Choice}
\author{Afonso Rodrigues\setfoot{unia}{Düsseldorf Institute for Competition Economics, Heinrich Heine University Düsseldorf (afboro@dice.hhu.de). A portion of this work was carried out during the author's doctoral studies at the University of Oxford. There are no relevant or material financial interests that relate to the research described in this paper.}}

\date{\today}

\begin{document}
\maketitle
\small

\begin{center}
    \textbf{Abstract}\\
\end{center}

Aggregate linear demand systems can be micro-founded by a population of heterogeneous consumers making discrete choices. They can also be extended to incorporate income effects. I prove these results and show how to structurally estimate this class of models using only market-level data. The distribution of heterogeneous consumer preferences is recovered by disciplining the corner solutions with aggregate terms only. This approach allows for the same richness of Berry, Levinsohn, and Pakes (1995) [BLP] while retaining the analytical and computational advantages of linear demand: it does not require exogenously fixing market size, allows for heterogeneity in choice sets, and results in a flexible substitution matrix which varies endogenously as different consumers enter/exit the market. The discrete-choice microfoundation facilitates welfare analysis and the flexible measurement of markups and cost pass-throughs, among a wide range of other applications.\\



\normalsize
%\doublespacing
\newpage
\section{Introduction}\label{introduction}

The main contribution of this paper is to investigate the conditions under which an aggregate linear demand system for differentiated goods can be derived from a population of heterogeneous consumers making discrete choices.\footnote{McFadden (1974), Daly and Zachary (1978), McFadden and Train (2000), and Anderson, de Palma, and Thisse (1989) do so for the multinomial, nested, and mixed logit, and Constant Elasticity of Substitution (CES) utility respectively.} More broadly, the paper shows that the tractability of linear demand need not come at the expense of heterogeneous consumer preferences, flexible substitution patterns, or economically meaningful welfare and policy counterfactuals.\\

Producing a discrete-choice microfoundation for linear demand is not in of itself novel. Hotelling (1929), Salop (1979), Chen and Riordan (2007), and Bos and Vermeulen (2022) provide settings in which discrete consumer choice generates linear or piecewise-linear demand: consumers have a one-dimensional distribution of valuation differences and can only choose between pairs of alternatives, occasionally including the outside good. These are remarkably restrictive kinds of consumer heterogeneity which impose tight conditions, particularly on cross-price effects. The literature suggests this may be unavoidable: Jaffe and Weyl (2010) show that linear demand cannot arise from discrete choice if we assume consumer valuations have a smooth, full-dimensional density across products; Armstrong and Vickers (2015) furthermore confirm linear demand requires some restriction on the support of consumer valuations.\footnote{It should be noted that Armstrong and Vickers (2015) already establish the necessary and sufficient conditions for a minimal linear demand system to be microfounded by discrete choice. As explained below, the present paper builds on this to produce a particular operationalisation of their result which allows for full-dimensional consumer valuations on aggregate, and does not require quasilinearity.}\\

As a result, these models have remained primarily a tool for theoretical research. In empirical work, aggregate linear demand is more often represented via a representative consumer with quasilinear quadratic utility (Amir, Erickson, and Jin, 2017; Choné and Linnemer, 2020). Examples include Melitz and Ottaviano (2008) in international trade; and Pinkse, Slade, and Brett (2002), Thomassen, Smith, Seiler, and Schiraldi (2017), and Pellegrino (2025) in industrial organisation. However, while allowing full flexibility in cross-price effects, the model cannot incorporate income effects, produces inflexible cost pass-throughs, and assumes homogeneity in consumer preferences, making it impractical for policy assessments and analyses.\\

The main novelty of this paper is therefore primarily \textit{generalisability}; to produce a discrete-choice microfoundation that can yield a fully flexible demand system which may be structurally estimated without a representative consumer assumption. I show this can be achieved by choice of a suitable distribution for the idiosyncrasies in tastes across individual consumers: aggregate linear demand arises when consumers are uniformly distributed over how much weight they place on a personal (cardinal) ranking of the products relative to their characteristics. Once aggregated, heterogeneity in the one-dimensional orderings allows rich substitution patterns mimicking full-dimensional valuations.\\

Income effects appear if the marginal utility of income across consumers is locally constant and differs in slope at the relevant pre- and post-purchase consumption levels. I show these can then be modelled as an intercept in the matrix of product characteristics inside the aggregate demand function. Incorporating income effects allows the linear demand system to operate as an incomplete demand system (LaFrance, 1985) while remaining fully consistent with utility maximization theory. This holds nice properties in settings where researchers wish to analyse competition amongst a subset of an entire assortment of available alternatives.\\

Consumer heterogeneity follows from the discrete choice microfoundation, but implies heterogeneous corner solutions. Discussions of corner solutions in aggregate linear demand are often neglected in favour of assuming an interior Nash equilibrium (Choné and Linnemer, 2020), removing any role for consumer heterogeneity. Different individual consumers reaching their corner solution - e.g. as price increases - produce kinks at different points in the aggregate demand function; without corners, there are no kinks, and the aggregate parameters are at best a (weighted) mean of individual parameters.\\

To address this, I introduce a novel tool to model corner solutions directly, which allows aggregate linear demand estimation to recover the full distribution of consumer preferences. It relies on the set of possible \textit{ordinal} orderings, or permutations, of the set of goods. I demonstrate how the full model can be disciplined using only aggregate terms by sketching out a methodology for the structural estimation of the aggregate demand system using market-level data exclusively. The methodology follows a step-by-step process: (1) estimate an initial unconstrained linear demand system using excluded instruments; (2) decompose the unconstrained demand Jacobian into simulated one-dimensional permutations, whose average reproduces the estimated unconstrained system exactly; (3) introduce a distribution of heterogeneity in characteristic preferences; and (4) solve the implied corner problem for each simulated permutation, aggregate the resulting corner corrections, and jointly re-estimate the constrained demand and heterogeneity parameters by simulated GMM.\\

The framework retains much of the flexibility of discrete-choice models such as BLP, while preserving the analytical and computational advantages of linear demand. Three features are particularly relevant, in that they follow directly from the model without additional complications. First, unlike conventional discrete-choice models, estimation does not require an exogenous definition of market size. Provided that within the observed price region there is always some potential consumers which do not buy, aggregate demand and its derivatives do not depend on the size of the underlying consumer population. Second, the framework accommodates heterogeneity in consumers' choice sets: different consumers may have access to different subsets of products, with these restrictions incorporated directly into their individual demand response matrices and, consequently, aggregate demand. Third, while unconstrained demand is linear, the aggregate substitution matrix need not remain constant as prices change. As different consumers reach or leave product-level corner solutions, the set of alternatives over which they substitute changes, endogenously altering the aggregate demand Jacobian. This allows for rich substitution patterns which vary with prices and consumer participation, without sacrificing the tractability of the underlying linear demand system.\\

The model shows potential for a wide range of applications, including where linear demand has previously not seen significant use. One common concern is the analysis of consumer welfare, particularly due to the possibility that counterfactual welfare changes depend on the path taken through the price vector. I show how the discrete-choice microfoundation enables welfare analysis by providing a corresponding expected-surplus representation. A second common concern is that linear demand implies constant cost pass-through: conditional on product characteristics and firm ownership, equilibrium pass-through is usually invariant to the level of prices and marginal costs. Corner solutions relax this restriction by introducing endogenous demand curvature. As prices change, different consumers reach product-level zero-demand constraints, changing the active demand Jacobian and therefore the degree of cost pass-through. This allows pass-throughs to depend on price and marginal cost, and rise beyond a half or even above one. The model's properties may also be valuable in settings where demand and equilibrium prices must be evaluated repeatedly, such as models of endogenous entry and exit and durable-goods competition. They can facilitate the analysis of large choice sets in retail markets, network-based substitution structures, and endogenous participation in platform markets.

\section{Main result}\label{main-result}

\subsection{Setup} 

Let $\mathcal{N}={0,1,\ldots,N}$ be the set of inside goods and the outside option, where $i=0$ is the outside option. Let $X\in\mathbb{R}^{N\times K}$ be the matrix of observed product characteristics, and let $\boldsymbol{p}\in\mathbb{R}^N$ be the vector of inside-good prices.\\

A set of consumers indexed $h\in\mathcal{H}$ chooses exactly one alternative from $\mathcal{N}$. For an individual consumer $h$, the indirect utility obtained from inside good $i$ is:
\begin{equation}
u_{hi}=\underbrace{\boldsymbol{x}_i'\boldsymbol{\beta}_h+{\xi}_i-p_i}_{\equiv \nu_{hi}\text{, deterministic inside-good utility}}+\varepsilon_{hi}
\end{equation}

where $\boldsymbol{\beta}_h\in\mathbb{R}^K$ is the consumer-specific vector of preference coefficients for the observed characteristics, $\boldsymbol{\xi}\in\mathbb{R}^N$ represents any latent utility for unobserved characteristics or demand shocks, and $\varepsilon_{hi}$ is stochastic utility. Normalise the utility of the outside option ($\nu_{h0}=0$) and let the stochastic utility distribution to follow a general form:
\begin{equation}
\boldsymbol{\varepsilon}_{h}
=
(\varepsilon_{h0},\ldots,\varepsilon_{hN})'\sim
F_h(\boldsymbol{\varepsilon}\mid X,\boldsymbol{\xi})
\end{equation}

The $X$ and $\boldsymbol{\xi}$ may determine, for example, the dispersion of stochastic utility, their correlation across products, or the extent to which consumers distinguish between similar products. The dependence of $F_h$ on $X$ and $\boldsymbol{\xi}$ furthermore permits $\boldsymbol{\varepsilon}_{h}$ to change with $\mathcal{N}$: e.g., stochastic utility for two products may be more strongly correlated when the products have similar characteristics.\\

Consumers choose whichever alternative yields them the highest indirect utility. Because this depends on the stochastic $\boldsymbol{\varepsilon}_h$, this decision rule pins down choice probabilities $\boldsymbol{z}_{h}$:
\begin{equation}\nonumber
i^*_{h}=\arg\max\{\max_{i\in\mathcal{N}}\{\nu_{hi}+\varepsilon_{hi}\},0\}
\end{equation}

The resulting choice probabilities for consumer $h$ are:
\begin{equation}\label{optprob}
z_{hi}=\Pr[i=\arg\max\{\max_{i\in\mathcal{N}}\{\nu_{hi}+\varepsilon_{hi}\},0\}\mid\boldsymbol{\nu}_h]
\end{equation}

\subsection{Perturbed utility representation}

Now, consider a candidate inside-good choice probabilities vector $\boldsymbol{z}$ for consumer $h$ with the usual properties:
\begin{equation}
\boldsymbol{z}=(z_1,\ldots,z_N)':\ \sum_{i=1}^Nz_i\leq1,z_i\geq0\ \forall i=1,\ldots N
\end{equation}

If a mass of consumers had the exact same preferences as consumer $h$, each $z_i$ could be seen as the fraction of these which would choose inside good $i$ based on their $\boldsymbol{\varepsilon}$ draws from $F_h$. In this case, $z_0=1-\boldsymbol{1}'\boldsymbol{z}$ would be the fraction choosing the outside option.\\

Another way to view this is that consumers are effectively \textit{sorting} themselves into whichever alternative yields them the highest indirect utility - \textit{as if} a "social planner" were maximising a group-level choice problem, allocating consumers to goods based on individual stochastic utility and the observed choice probabilities. This could not be just any allocation. For a given candidate $\boldsymbol{z}$, consumers should be assigned to each product in order to make those shares consistent with the greatest possible aggregate stochastic utility; otherwise, the decision of the group-level "social planner" and that of the individual consumers would differ. A "social planner" would need to work backwards so that, conditional on an observed candidate $\boldsymbol{z}$ and given the distribution of stochastic errors, it may identify which consumers most likely bought each good such that it delivers $\boldsymbol{z}$.\\

This is the intuition behind the following result. Let $\mathcal{P}_h(\boldsymbol{z})$ be a possible collection of partitions $\{A_0,A_1,\ldots,A_N\}$ of the stochastic-utility space of consumers with preferences identical to those of consumer $h$, satisfying:
\begin{equation}
F_h(A_i\mid X,\boldsymbol{\xi})=\int_{A_i}dF_h(\boldsymbol{\varepsilon}|X,\boldsymbol{\xi})\ =z_i\qquad F_h(A_0\mid X,\boldsymbol{\xi})=1-\boldsymbol{1}'\boldsymbol{z}\qquad\ \forall i=1,\ldots,N
\end{equation}

The set $A_i$ represents one possible set of these consumers assigned to alternative $i$ given our assignment rule. Unit demand and discrete choice require the size of each set $A_i$ to equal the corresponding choice share. The candidate $\boldsymbol{z}$ enters the assignment problem by determining the size of each subset of this group of otherwise identical consumers assigned to each product. In this sense, $F_h$ provides the correspondence between the observed market share of each good to the set of specific consumers which bought it. Think of the $\boldsymbol{\varepsilon}$ of each consumer as different pieces of cake. Consumers are partitioned based on what size piece they received for each good, and keep only the piece of cake corresponding to the good of their partition. Then, the total size of all pieces in each partition of each good must be equivalent to the consumer-level choice probability of that good.\\

To obtain the collection of partitions which correspond to a given candidate $\boldsymbol{z}$, define:
\begin{equation}
\mathcal{Z}_h(\boldsymbol{z},X,\boldsymbol{\xi})=\sup_{{A_i}_{i=0}^N\in\mathcal{P}_h(\boldsymbol{z})}\{\sum_{i=0}^N\int_{A_i}\varepsilon_i\ dF_h(\boldsymbol{\varepsilon}\mid X,\boldsymbol{\xi})\}
\end{equation}

The function $\mathcal{Z}_h(\boldsymbol{z},X,\boldsymbol{\xi})$ is the greatest aggregate stochastic utility attainable while delivering the choice shares specified by $\boldsymbol{z}$. The maximising partition assigns each product to those consumers for whom that product has the strongest stochastic advantage relative to the alternatives. This structure follows from transport theory, and is equivalent to the allocation of a continuously distributed mass - in this case, of consumers - to finitely many destinations - in this case, products - subject to capacity/mass constraints - in this case, product shares (Galichon and Salanié, 2022; see also Rockafellar, 1970; Hofbauer and Sandholm, 2002; Fosgerau, Melo, de Palma, and Shum, 2020).\\

\textit{\textbf{Proposition 1}: The consumer-level choice probabilities for consumer $h$ admit the equivalent perturbed utility representation:}
\begin{equation}\label{segmentleveloptimisation}\nonumber
\boldsymbol{z}_h(\boldsymbol{\nu}_h)=\arg\max_{\boldsymbol{z}:\sum_{i=1}^Nz_i\leq1,z_i\geq0\ \forall i=1,\ldots N}
\{\boldsymbol{\nu}_h'\boldsymbol{z}+\mathcal{Z}_h(\boldsymbol{z},X,\boldsymbol{\xi})\}
\end{equation}

For each candidate $\boldsymbol{z}$, the term $\boldsymbol{v}'\boldsymbol{z}$ records the systematic utility associated with those shares, while $\mathcal{Z}_h(\boldsymbol{z},X,\boldsymbol{\xi})$ records the greatest stochastic utility attainable by sorting consumers across products in a manner consistent with those shares.\\

% Suppose that a fraction $z_i$ of the segment must choose good $i$. The optimal assignment first places in $A_i$ those consumers with particularly high values of $\varepsilon_i$ relative to their stochastic utility for other goods. If $z_i$ is increased, the set $A_i$ must expand. The additional consumers assigned to good $i$ will generally have weaker comparative preferences for it than those already assigned. The stochastic quality of the marginal match therefore declines as the required share of good $i$ increases. This sorting logic also operates across products. Increasing $z_j$ changes which consumers remain available to be assigned to good $i$. If goods $i$ and $j$ appeal to similar consumers, expanding the share of $j$ may remove from the remaining population precisely those consumers who would otherwise have been strong matches for $i$.\\

\subsection{Linear demand}

In this paper, we consider the following question. Under what conditions does a population of heterogeneous consumers making discrete choices yield aggregate linear demand across an arbitrary number of differentiated goods?\\

We can rely on the concavity of $\mathcal{Z}_h(\boldsymbol{z},X,\boldsymbol{\xi})$ to answer this question.\footnote{Consider two feasible share vectors $\boldsymbol{z}$ and $\boldsymbol{\widetilde z}$, together with assignments attaining their corresponding values of $\mathcal{Z}_h$. For any $\lambda\in[0,1]$, apply the first assignment to a fraction $\lambda$ of consumers identical to consumer $h$ and the second assignment to the remaining fraction $1-\lambda$. This produces the combined shares $\lambda \boldsymbol{z}+(1-\lambda)\boldsymbol{\widetilde z}$ and delivers aggregate stochastic utility $\lambda \mathcal{Z}_h(\boldsymbol{z},X,\boldsymbol{\xi}) +(1-\lambda)\mathcal{Z}_h(\boldsymbol{\widetilde z},X,\boldsymbol{\xi})$. Because $\mathcal{Z}_h$ selects the best assignment consistent with the combined shares, it must satisfy $\mathcal{Z}_h(\lambda \boldsymbol{z}+(1-\lambda)\boldsymbol{\widetilde z},
X,\boldsymbol{\xi})\geq\lambda\mathcal{Z}_h(\boldsymbol{z},X,\boldsymbol{\xi})
+(1-\lambda)\mathcal{Z}_h(\boldsymbol{\widetilde z},X,\boldsymbol{\xi})$.}  If $\mathcal{Z}_h(\boldsymbol{z},X,\boldsymbol{\xi})$ were quadratic in $\boldsymbol{z}$, (\ref{segmentleveloptimisation}) would resemble quadratic utility, commonly used to reproduce linear demand from a representative consumer's preferences (see Amir, Erickson, and Jin, 2017; and Choné and Linnemer, 2020). The following class of stochastic utility distributions makes a $\mathcal{Z}_h(\boldsymbol{z},X,\boldsymbol{\xi})$ quadratic in $\boldsymbol{z}$ possible:
\begin{equation}\label{thealphas1}
\forall i,h:\quad\varepsilon_{hi}(X,\boldsymbol{\xi})=\alpha_{hi}(X,\boldsymbol{\xi})\epsilon_{h}\quad\epsilon_{h}\sim U[0,1]
\end{equation}
where each consumer-and-product-specific $\alpha_{hi}$ can be represented within a Hotelling-style one-dimensional ordering:
\begin{equation}\label{thealphas2}
    \alpha_{hi_1}<\alpha_{hi_2}<\ldots<\alpha_{hi_N}<\alpha_0=0
\end{equation}

% \footnote{This is equivalent to assuming \textit{local substitution} for a particular product category with \textit{single-crossing preferences} - i.e., as the consumer-specific stochastic preference term rises, the utility-maximising alternative for that consumer "moves through" the products according to their ordering on the $\alpha$-line a single time; and \textit{uniform latent heterogeneity}.}

This defines $\boldsymbol{\varepsilon}$ as a penalty on deterministic utility. The $\alpha$ terms are product-specific with unspecified dependence on $X$. On the one hand, a product-specific adjustment means that two goods with the same observed characteristics need not have identical indirect utility; ont he other hand, $\ref{thealphas1}$ can be understood as a possibly non-linear correction to how products derive their utility from the characteristics they hold.\\

Under this distribution, the optimal assignment of identical consumers with different stochastic utility to products yields the following:
\begin{equation}
\begin{aligned}
\mathcal Z_h(\boldsymbol z,X,\boldsymbol\xi)
&=
\sup_{{A_i}\in\mathcal P_h(\boldsymbol z)}
\sum_{i=0}^N\int_{A_i}
\alpha_{hi}(X,\boldsymbol{\xi})\epsilon\ d\epsilon
\\
&=
\sum_{i=1}^N
\alpha_{hi}
\int_{\sum_{j:\alpha_{hj}<\alpha_{hi}}z_j}^{\sum_{j:\alpha_{hj}\leq\alpha_{hi}}z_j}
\epsilon\ d\epsilon
\\
&=
-\frac12\sum_{i=1}^N|\alpha_{hi}|z_i^2
-
\sum_{\substack{i<j}}
\min\{|\alpha_{hi}|,|\alpha_{hj}|\}z_i z_j
\\
&=
-\frac12
\sum_{i=1}^N\sum_{j=1}^N
z_i z_j
[
\frac{|\alpha_{hi}|+|\alpha_{hj}|-|\alpha_{hi}-\alpha_{hj}|}{2}
]
\\
&=
-\frac12\boldsymbol z'M_h(X,\boldsymbol\xi)\boldsymbol z,\qquad \text{for}\ \{M_h\}_{ij}
=
\frac{
|\alpha_{hi}|+|\alpha_{hj}|-|\alpha_{hi}-\alpha_{hj}|
}{2}
\end{aligned}
\end{equation}

The element $\{M_h\}_{ij}$ has a simple geometric interpretation: it is the length of the
overlap between the intervals joining $0$ to $\alpha_{hi}$ and $0$ to $\alpha_{hj}$ on the consumer's one-dimensional ordering. The consumer choice problem therefore becomes exactly:
\begin{equation}\label{seglevelchoiceprob}
\boldsymbol{z}_h=\arg\max_{\boldsymbol{z}\geq\boldsymbol{0},\ \mathbf{1}'\boldsymbol{z}\leq1}
\left\{\boldsymbol{\nu}_h'\boldsymbol{z}-\frac{1}{2}\boldsymbol{z}'M_h\boldsymbol{z}\right\}=M_h^{-1}(X\boldsymbol{\beta}_h+\boldsymbol{\xi}-\boldsymbol{p})
\end{equation}

The latter equality representing the unconstrained case, where the probability constraints are slack. We cover the general case below.\\

That truncated linear choice probabilities follow from a random-utility model with uniformly distributed stochastic utility has been known since at least Domencich and McFadden (1975); the concept nests Hotelling (1929) much earlier. Nonetheless, two novelties stand out. Firstly, this framing allows us to extend the concept to a setting with multiple goods. The one-dimensional representation of the cardinal ranking is visibly imposing structure on $M_h^{-1}$ of any individual consumer. Let $d_{h,i_ki_{k+1}}=|\alpha_{hi_{k+1}}-\alpha_{hi_k}|$ be the distance between consecutive alternatives. Given $M_h$, the respective response matrix $L_h$, which includes the outside option, is as follows:
\begin{equation}\label{responsematrixx}
L_h=\begin{pmatrix}
\mathbf{1}'M_h^{-1}\mathbf{1}
&-(M_h^{-1}\mathbf{1})'\\
-M_h^{-1}\mathbf{1}&M_h^{-1}
\end{pmatrix}
=
\sum_{k=0}^{N-1}
\frac{1}{d_{h,i_ki_{k+1}}}
(e_{i_k}-e_{i_{k+1}})
(e_{i_k}-e_{i_{k+1}})'
\end{equation}

The $\alpha$ terms allow for greater complexity in the distribution of consumer preferences - after all, $u_{hi}$ need not be linear in characteristics. Furthermore, note what this implies for the necessary $\varepsilon_{hi}$ for consumer $h$ to prefer good $i$ to good $j$:

\begin{equation}
    \begin{aligned}
        \nu_{hi}+\alpha_{hi}\epsilon^{*_{h,ij}}
        =\nu_{hj}+\alpha_{hj}\epsilon^{*_{h,ij}}
        \quad\Leftrightarrow\quad
        \epsilon^{*_{h,ij}}=\frac{\nu_{hi}-\nu_{hj}}{\alpha_{hj}-\alpha_{hi}}\in[0,1]
    \end{aligned}
\end{equation}

The threshold $\epsilon^{*_{h,ij}}$ is increasing in the difference between the deterministic utilities of the two goods - meaning that the switch is less likely to happen - and decreasing in the difference between the penalty terms. One may think that this means that similar goods are (paradoxically) less likely to be close substitutes, but the model suggests something more interesting: an incentive towards differentiation on the one hand, as can be seen here, combined with consumer-level \textit{local substitution}.\\

Consumers are said to exhibit local substitution if, when a preferred car becomes unavailable or sufficiently unattractive, substitution is concentrated on the nearest available alternative in that ordering (for evidence of this in e.g. the automobile industry, see Berry, Levinsohn, and Pakes, 2004, and Grieco, Murry, and Yurukoglu, 2024). This is exactly what we observe in Figure 1. Figure 1 provides a useful two-good, two-consumer illustration of this section's core result. The consumers differ both in preferences for characteristics and how they order alternatives outside those preferences, represented by the x-axis. "Segment" refers to consumers identical to consumer $h$ but with different $\epsilon$ draws. The y-axis is the total indirect utility of each alternative. At higher $\varepsilon_{h}$ we have consumers where the cardinal ordering holds greater weight - this can be sufficient to upend the initial ordering for a consumer, as produced by deterministic utility. On the left, this means a consumer going from preferring good 2 (in green), to good 1 (in yellow), to finally preferring to not buy anything at all (the outside option, in blue). On the right, consumers start preferring good 1, then good 2, and finally the outside option. These movements along the y-axis produce these alternatives' respective choice probabilities for each consumer - the bars observed at the bottom of each of these sub-figures. The cardinal ordering can be perceived as a penalty of varying strengths on the deterministic utility derived from product characteristics, shifting indirect utility orderings without loss of generality.\\

The sub-figure on the bottom reproduces what aggregate demand looks like for a population composed of 45\% consumers like that on the left graph, and 55\% consumers like that on the right graph, with a kink at the reservation price of the first group. The price increase offers an additional dimension. Notice how when price of good 2 increases, only the market share of good 1 changes in the sub-figure on the left - reflecting the idea of local substitution. Consumers of good 2 in the left sub-figure only move to the outside option after rejecting both good 2 and good 1. The same does not apply in the sub-figure on the right, where we see substitution to good 1 and the outside option. \\

\begin{figure}[H]
\label{fig:placeholder}
    \centering
    \caption{Two-good, two-consumer example under a price increase}
    \includegraphics[width=\linewidth]{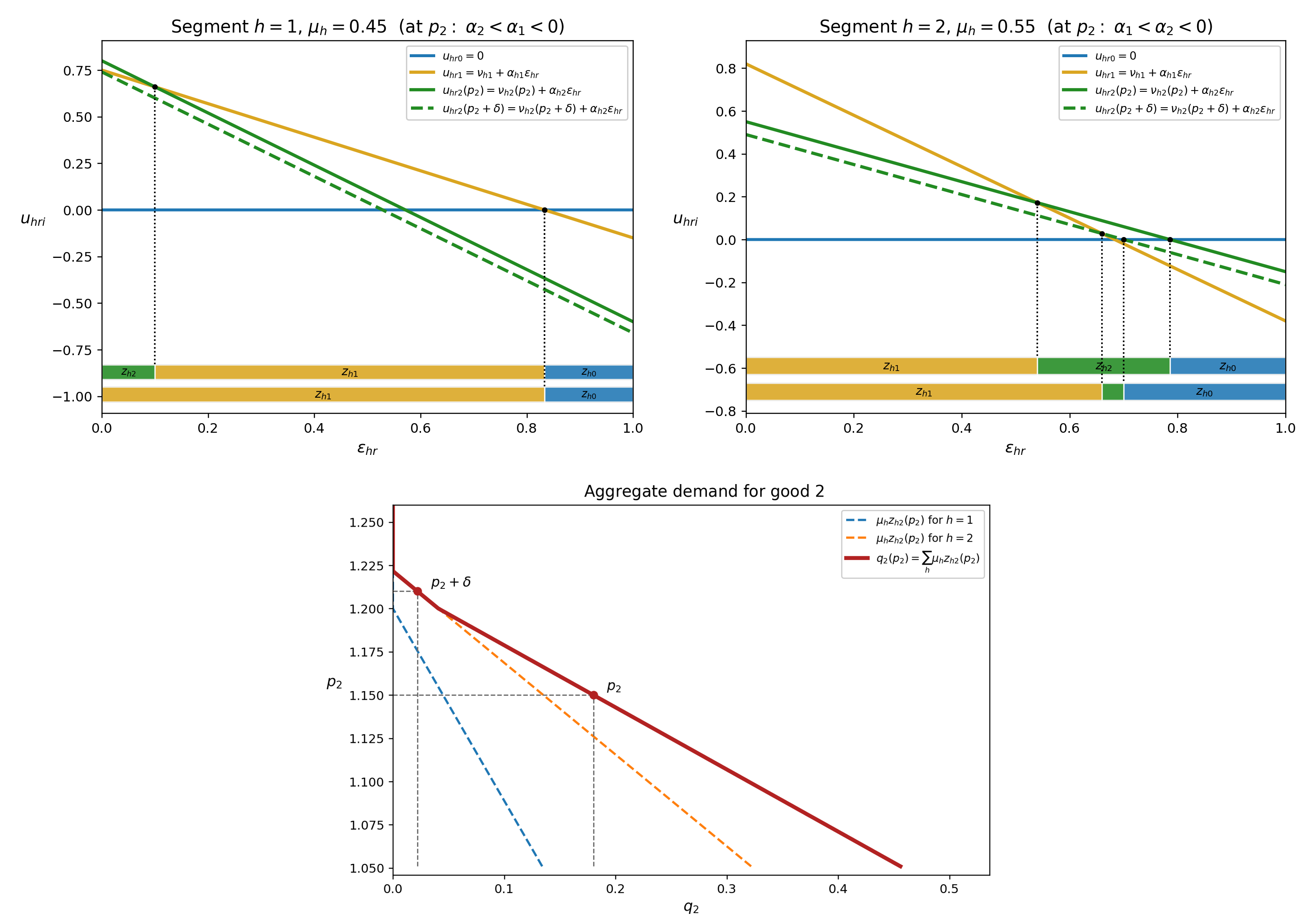}
\end{figure}

For a particular consumer, alternatives are only substitutes when they are adjacent on that consumer's one-dimensional ordering - confirming local substitution. Altogether, greater similarity in $\alpha$ does mean a lower likelihood that a consumer goes from preferring one good to preferring another, but \textit{conditional} on these being direct competitors. Should one of the goods exit the market, it would be to this alternative that much of the other's market share would move to.\\

As for the second novelty: once we aggregate across consumers, this consumer-level one-dimensionality in preferences with pairwise competition between adjacent alternatives disappears; instead, the model delivers flexible, multi-dimensional substitution. This is because different consumers may order alternatives differently, so different pairs of alternatives will be adjacent for different consumers. The following properties survive aggregation, however:

\begin{equation}\label{conditionsagg_1}
    \begin{aligned}
        \{M_h^{-1}\}_{ij}&=\{L_h\}_{ij}=\begin{cases}
        -\frac{1}{|\alpha_{hi}-\alpha_{hj}|}&\text{if $i,j$ are consecutive}\\
        0&\text{otherwise}
    \end{cases}\\
    \Leftrightarrow\{M_h^{-1}\}_{ij}&\leq0,\ \forall i\neq j
    \end{aligned}
\end{equation}
and
\begin{equation}\label{conditionsagg_2}
\begin{aligned}
        &L_h\boldsymbol{1}_{N+1}=0=\{L_h\}_{i0}+\sum_{j=1}^N\{M_h^{-1}\}_{ij}\\
        \Rightarrow\ &\{M_h^{-1}\boldsymbol{1}\}_i=-\{L_h\}_{i0}=\begin{cases}
        \frac{1}{|\alpha_{hi}-\alpha_0|}&\text{if $i,0$ are consecutive}\\
        0&\text{otherwise}
    \end{cases}\geq0,\ \forall i\\
    \Leftrightarrow\ & M_h^{-1}\boldsymbol{1}\geq0
\end{aligned}
\end{equation}

To see this, let $M^{-1} = \sum_{h\in\mathcal{H}} M_h^{-1}$. Then, we have:
\begin{equation}\label{aggregate_conditions}
\begin{aligned}
[M^{-1}]_{ij}
&=
\int_{\mathcal H}\{M_h^{-1}\}_{ij}\ dh
\leq0,
\ \forall i\neq j,\qquad
M^{-1}\boldsymbol 1
=
\int_{\mathcal H}M_h^{-1}\boldsymbol 1\ dh
\geq0
\end{aligned}
\end{equation}

% The importance of this is clear. We have shown that any symmetric positive-definite $M^{-1}$ satisfying (\ref{conditionsagg_1}) and (\ref{conditionsagg_2}) can be generated exactly by a finite mixture of these globally quadratic one-dimensional random-utility segments. In Section \ref{accounting-for-corner-solutions}, I show how this aggregate term can be used to discipline the sum of individual corner corrections, needed once we impose non-negativity constraints on consumer-level choice probabilities. In Section \ref{network-theory}, I explore the aggregate demand implications of some well-known network graphs, where $M$ is interpreted as an adjacency matrix.\\

This completes our discussion on how individual consumer choice probabilities can be framed via a linear demand system. Under what conditions can we aggregate across consumers to obtain aggregate linear demand?\\

Aggregate demand is:
\begin{equation}
q(\boldsymbol{p},X,\boldsymbol{\xi})=\sum_{h\in\mathcal{H}}
\mu_hM_h^{-1}(X\boldsymbol{\beta}_h+\boldsymbol{\xi}-\boldsymbol{p}).
\end{equation}

Suppose characteristic preferences satisfy $\mathbb{E}[\boldsymbol{\beta}_h|\boldsymbol{\xi}]=\boldsymbol\beta$. Then, it follows that:
\begin{equation}\label{baselineaggregate}
\boldsymbol{q}(\boldsymbol{p},X,\boldsymbol{\xi})=\left[
\sum_{h\in\mathcal{H}}\mu_hM_h^{-1}
\right]
(X\boldsymbol{\beta}+\boldsymbol{\xi}-\boldsymbol{p})=M^{-1}(X\boldsymbol{\beta}+\boldsymbol{\xi}-\boldsymbol{p})
\end{equation}
for $M^{-1}=\sum_{h\in\mathcal{H}}\mu_hM_h^{-1}$.\\

Note that $\boldsymbol{q}(\boldsymbol{p},X,\boldsymbol{\xi})$ is not a vector of probabilities, but a consumption vector. This not only directly ties our demand system to that usually obtained from representative consumer models, but allows us to avoid making assumptions regarding market size. This demand system requires neither the exogenous definition of market size nor a separate market-size model. We obtain this by working from the assumption that $\boldsymbol{1}'\boldsymbol{z}<1$, meaning that there is a non-negligible number of consumers who decide not to buy at any price in its support (e.g. across all prices in a dataset). We keep this assumption throughout. This is implicitly assumed in logit models as well. In our setting, however, this is enough that different latent population sizes can produce exactly the same demand function. Population size is then redundant for estimating that function and its derivatives. In logit models, however, market size is not just a scaling term, but affects e.g. the price derivatives and elasticities themselves. What market size we choose can alter these elasticities, therefore producing bias in their estimation.\\

The model can incorporate heterogeneous choice sets across consumers. To see this, let $\mathcal J_h$ be the set of inside goods available to consumer $h$. A good $j\notin\mathcal J_h$ is therefore structurally unavailable to that consumer and has zero choice probability at every price; it is not at a price-induced corner. The set $\mathcal C_h(\boldsymbol{p},X)\subseteq\mathcal J_h$ contains only corners among goods which the consumer could - in principle - purchase; permanent choice-set exclusions can be directly incorporated in the aggregate $M^{-1}$ via individual $M^{1}_h$. For $S_h$ the $N\times |\mathcal J_h|$ selection matrix which matches the goods in $\mathcal J_h$ into the full product space, and starting from $\widetilde M_h^{-1}$ as the $|\mathcal{J}_h|\times |\mathcal{J}_h|$ unconstrained demand Jacobian for consumer $h$, we have:
\begin{equation}
    M_h^{-1}\equiv S_h\widetilde M_h^{-1}S_h'\quad\Rightarrow\quad M^{-1}= \int S_h\widetilde M_h^{-1}S_h'\,d\mu(h)
\end{equation}

We explore how this matrix architecture can used to allow for discrete choice models over baskets of goods - and how this may be an avenue for introducing complementarities between goods -  in Section \ref{retail}.\\

% We have not assumed the existence of a product-specific idiosyncratic term $\boldsymbol{\xi}\in\mathbb{R}^N$, which is commonly used in discrete-choice models to refer to unobserved product quality or a product-market demand shock; this is WLOG if it varies only by product. We allow its inclusion in the aggregate linear demand function going forward, and generalise the characteristics preference assumption $\mathbb E[\boldsymbol{\beta}_h]=\boldsymbol{\beta}$\footnote{The assumption does not require, in any of its forms, that $\boldsymbol{\beta}_h=\boldsymbol{\beta}$ for every consumer; it requires only that the conditional mean of characteristic preferences does not vary systematically across each consumer's demand.}.\\

% We have worked throughout with consumer segments. Albeit useful, the segmentation of consumers is not necessary for any of the arguments in this section or the paper. Rather than consumer-level choice probabilities, $\boldsymbol{z}_h$ could instead be a consumer-level probability of choosing amongst available alternatives prior to drawing $\varepsilon_h$. While we have separated the consumer-level $\alpha$ orderings from consumer-level $\epsilon$ throughout this section, both can be consumer-specific in practice. All points would then follow through via expected value. Nonetheless, because discussing concepts like $\mathcal{Z}$ can be less intuitive without segments, we retain the segmentation rationale for the rest of the paper.
\section{Extensions}
\subsection{Income effects}\label{income-effects}

In this section, I introduce income effects to aggregate linear demand. This is not so much a problem in models such as Hotelling (1929), Salop (1979), or Chen and Riordan (2007), yet as mentioned above the more common tool for linear demand estimation is quasilinear (i.e. income-effect-free) quadratic utility. In practice, this means that linear demand estimation with income effects is to my knowledge absent in the literature.\\

Allowing for income effects has an important implication: that linear demand systems may successfully operate \textit{incomplete demand system}, where demand for a set of inside goods can be analysed without specifying consumers' entire consumption vector (Epstein, 1982; LaFrance, 1985). The goal is to find general conditions under which we can include income into our discrete-choice microfoundation above. This is unlike the very restrictive conditions that the previous literature targeting broader integrability requires (LaFrance, 1985), as we focus instead on aggregation.\\

Let $y_{h}$ be the income of consumer $h$ and let $g(y_{h}-p_i)$ be their utility from residual income consumption. We are going to use the simplifying assumption that the stochastic utility term is distributed across consumers independently from income, i.e.:
\begin{equation}\label{Distrib_assump}
    F_h(\boldsymbol\varepsilon|y)=F_h(\boldsymbol\varepsilon)
\end{equation}

The utility from buying interior good $i$:
\begin{equation}
u_{hi}=x_i'\boldsymbol{\beta}_h+\xi_i+g(y_{h}-p_i)+\varepsilon_{hi}, \forall i=1,\ldots,N\qquad u_{h0}=g(y_{h})+\varepsilon_{h0}
\end{equation}

After normalising relative to the outside option, we have:
\begin{equation}
\nu_{hi}=x_i'\boldsymbol{\beta}_h+\xi_i+g(y_{h}-p_i)-g(y_{h})
\label{eq:nu-income}
\end{equation}

Now let $\Delta g(y_{h},\boldsymbol{p})=(g(y_{h}-p_1)-g(y_{h}),\ldots,g(y_{h}-p_N)-g(y_{h}))'$. From (\ref{seglevelchoiceprob}) and (\ref{Distrib_assump}), it follows that the consumer-level inside-good choice probabilities can be obtained conditional on the distribution of $y_{h}$. To see how, define $\boldsymbol{z}_h(y_{h})$ the consumer-$h$ choice probability with income $y_{h}$. Then we can write:
\begin{equation}
    \boldsymbol{z}_h(y_{h})=\arg\max_{\boldsymbol{z}\geq 0,\boldsymbol{1}'\boldsymbol{z}\leq 1}\{\boldsymbol{\nu}_h(y_{h})'\boldsymbol{z}-\frac{1}{2}\boldsymbol{z}'M_h\boldsymbol{z}\}=M_h^{-1}\boldsymbol{\nu}_h(y_{h})=M_h^{-1}(X\boldsymbol{\beta}_h+\boldsymbol{\xi}+\Delta g(y_{h},\boldsymbol{p}))
\end{equation}

with slack probability constraints. Lastly, aggregating across consumers, we obtain:
\begin{equation}
    \boldsymbol{q}(\boldsymbol{p},X)=\sum_h\mu_hM_h^{-1}(X\boldsymbol{\beta}_h+\boldsymbol{\xi}+\Delta g(y_h,\boldsymbol{p}))
\end{equation}

Under what additional conditions does the above expression simplify further into aggregate linear demand? One way to recover linear demand is to impose a simple local restriction on the utility from residual income. Suppose that, over the relevant range, $g(\cdot)$ is linear both around each consumer's pre-purchase income $y_{h}$ and over the range containing all post-purchase residual incomes $y_{h}-p_i$, with different slopes. Then, for consumer $h$:
\begin{equation}
    \begin{aligned}
        &\ g(y_{h}-p_i)-g(y_{h})=\underbrace{(a_{h}^1-a_{h}^0)+(\lambda_{h}^1-\lambda_{h}^0)y_{h}}_{\equiv\ \eta_{h}}-\lambda_{h}^1p_i,\quad \forall i=1,\ldots,N\\
        \Rightarrow&\ \Delta g(y_{h},\boldsymbol{p})=\eta_{h}\boldsymbol{1}-\lambda_{h}^1\boldsymbol{p}
    \end{aligned}
\end{equation}

Under a monotonically concave $g(\cdot)$, $\lambda^1\geq\lambda^0>0$. The intercept and slope of the difference can vary across consumers. By allowing $y_{h}$ and $y_{h}-p_i$ across all goods to belong to different sections of $g(\cdot)$ with different slope and intercept, these restrictions can either be taken literally or interpreted as a first-order approximation to $g(\cdot)$.\\ 

The difference in the slope of $g(\cdot)$ at the pre- and post-purchase residual income is what produces the income effect in indirect utility:
\begin{equation}
    \frac{\partial u_{hi}}{\partial p_i}=-\lambda_{h}^1\qquad\frac{\partial u_{hi}}{\partial y_{h}}=\lambda_{h}^1-\lambda_{h}^0\\
    \qquad\Rightarrow\qquad\frac{\partial u_{hi}}{\partial p_i}=-\lambda_{h}^0-\frac{\partial u_{hi}}{\partial y_{h}}
\end{equation}

This contrasts with the standard quasi-linear specification of utility, where a price change leaves the marginal utility of income constant (equal to $\lambda_{h}^0$) which yields no income effects.\\

With these functional form assumptions on $g(\cdot)$, we can re-write our consumer-level choice probabilities:
\begin{equation}
    \boldsymbol{z}_h=M^{-1}_h(X\boldsymbol{\beta}_h+\boldsymbol{\xi}+\Delta g(y_{h},\boldsymbol{p)})=M^{-1}_h(X\boldsymbol{\beta}_h+\boldsymbol{\xi}+\eta_h\boldsymbol{1}-\lambda_h^1\boldsymbol{p})
\end{equation}

Normalisations drop the $\lambda_h^1$ from $\boldsymbol{p}$ by correcting $M^{-1}_h,\ \boldsymbol{\beta}_h,$ and $\eta_h$ - we do so going forward without altering their notation.\\

Our last step is to assume:
\begin{equation}
    \mathbb{E}[(\boldsymbol{\beta}_h,\eta_{h},\lambda_h^1)]=(\boldsymbol{\beta},\eta,\lambda^1)
\end{equation}
which allows us to write our final expression for aggregate linear demand:
\begin{equation}
    \boldsymbol{q}(\boldsymbol{p},X)=M^{-1}(X\boldsymbol{\beta}+\boldsymbol{\xi}+\eta\boldsymbol{1}-\boldsymbol{p})
\end{equation}

Then:\\

\textit{\textbf{Proposition 2}: Under a} price increase, \textit{we can separate income from substitution effects in aggregate linear demand by decomposing the Marshallian price derivative using the Slutsky equation:}

\begin{equation}\label{slutskyeq}\small\nonumber
    \frac{\partial q(\boldsymbol{p},X,\boldsymbol{\xi},\boldsymbol{y)}}{\partial \boldsymbol{p}}=-M^{-1}=\underbrace{-\sum_h\mu_h[\lambda_h^1M_h^{-1}-(\lambda_h^1-\lambda_h^0)\cdot\mathrm{diag}(M_h^{-1}\boldsymbol{1})]}_{\text{substitution effect}}-\underbrace{\sum_h\mu_h(\lambda_h^1-\lambda_h^0)\cdot\mathrm{diag}(M_h^{-1}\boldsymbol{1})}_{\text{income effect}}
\end{equation}

The intuition here is that, outside substitution effects, price increases decrease residual income; they shift all inside goods relative to the outside option, and therefore only generate switching from inside-good to outside-good purchases. The consumers affected by this, those who shift their choice following the price increase, are current purchasers who decide not to purchase anymore. This means that the pre-change choice is for the same good whose inside-outside boundary is crossed, delivering $(\lambda_h^1-\lambda_h^0)\cdot\mathrm{diag}(M_h^{-1}\boldsymbol{1})$.\\

For completeness, a marginal price \textit{decrease} gives a different compensated derivative, and thus a different substitution effect. While with a price increase the consumers who switch their choice are current inside-good buyers moving to the outside option, with a price decrease we observe only the opposite move. This means that, at the pre-change choice, there is no income effect in that instance.\\

% $d_{hi}^M=0$, meaning $\mathbb{E}[(D_yd_{h}^M)d_{h}^{M'}|h]=0$, and thus $D_p^- z_h^H=-\lambda_h^1M_h^{-1}=D_pz_h^M$

Under the conditions which deliver aggregate linear demand, income enters $\boldsymbol{q}(\boldsymbol{p},X)$ as a market-level demand shock. Notably, we have produced income effects without creating Slutsky asymmetry, due to these operating exclusively through the inside–outside margin. This is significant, in that the model maintains the conditions required for welfare analysis with consumer surplus without path dependence (Tirole, 1988).\\

\subsection{Corner solutions}\label{corner-solutions}

% Purpose of this section is to establish the following results:
% 1) consumer-level demand is piecewise linear, and the KKT conditions give the exact correction to unconstrained demand.
% 2) Aggregating these corrections gives $\kappa(p,X)$. Unconstrained aggregate demand depends on the mean of $\beta_h$, while $\kappa$ generally depends on its distribution.
% 3) With sufficiently continuous heterogeneity, aggregate demand can be smooth despite individual segments having kinks.

Let us return to the pre-income-effects demand model above. Said aggregate linear demand system assumes that the probability constraints in each consumer's choice problem are slack. In this section, I relax the non-negativity constraints while retaining the same discrete-choice microfoundation. The goals are: (i) to characterise aggregate demand when some consumer product shares reach zero, and (ii) to formalise the idea that, with enough consumer heterogeneity, the resulting piecewise linearity can be smoothed over. While this means departing from aggregate linear demand in general, it is the linearity of the unconstrained demand that facilitates analysis of non-linear corner solutions, and, as we see later, constrains them sufficiently for structural estimation.\\

The results shared in this section are primarily of a technical nature. More empirically-minded researchers may prefer to skip this section entirely and move on to Section $\ref{accounting-for-corner-solutions}$, where we use the regularity properties discussed here to address corner solutions using only market-level data. \\

From (\ref{seglevelchoiceprob}), recall that the actual consumer-level choice probabilities can be derived as follows:
\begin{equation}
\boldsymbol{z}_h(\boldsymbol{\nu})
=
\arg\max_{\boldsymbol{z}\geq \boldsymbol{0},\ \boldsymbol{1}'\boldsymbol{z}\leq 1}
\left\{
\boldsymbol{\nu}'\boldsymbol{z}-\frac12 \boldsymbol{z}'M_h\boldsymbol{z}
\right\}
\end{equation}

To isolate product-level corner solutions, be reminded that throughout the paper the outside-option constraint remains slack for all prices in its support; that is to say, there is always a small set of consumers who do not purchase at any relevant price. The only relevant constraints are therefore $z_{hj}\geq0,\ \forall h, j$.\\

For any price vector, let $\mathcal C_h(\boldsymbol{\nu})\equiv\left\{i:z_{hi}(\boldsymbol{\nu})=0\right\}$ be the set of inside goods that are at a corner for consumer $h$, of size $|\mathcal{C}_h|$ dependent on price and product characteristics (references to those are dropped from the subscript). The set $\mathcal C_h$ may contain any number of products. The Karush-Kuhn-Tucker conditions are:
\begin{equation}\label{exex}
M_h\boldsymbol{z}_h=X\boldsymbol{\beta}_h+\boldsymbol{\xi}-\boldsymbol{p}+\boldsymbol{\lambda}_h,\quad \lambda_{hj}\geq0,\quad  z_{hj}\geq0,\quad \lambda_{hj}z_{hj}=0,\ \forall j
\end{equation}

for $\boldsymbol{\lambda}_h\geq0$ be the vector of multipliers on the non-negativity constraints. The binding constraints require:
\begin{equation}
0=\{\boldsymbol{z}^I_{h}\}_{|\mathcal{C}_h|\times 1}+\{M^{-1}_{h}\}_{|\mathcal{C}_h|\times |\mathcal{C}_h|}\{\boldsymbol{\lambda}_h\}_{|\mathcal{C}_h|\times 1}\quad\Rightarrow\quad
\{\boldsymbol{\lambda}_h\}_{|\mathcal{C}_h|\times 1}=-(\{M_{h}^{-1}\}_{|\mathcal{C}_h|\times |\mathcal{C}_h|})^{-1}\{\boldsymbol{z}^I_{h}\}_{|\mathcal{C}_h|\times 1}
\end{equation}

for $\{\cdot\}_{|\mathcal{C}_h|\times |\mathcal{C}_h|}$ and $\{\cdot\}_{|\mathcal{C}_h|\times 1}$ operators which select the rows and columns indexed by those goods in $\mathcal C_h$ from the object they are related to. The subscripts therefore contain information on the dimensions of matrices and vectors impacted by the set of inside goods at a corner.\\

Consumer $h$'s choice probabilities are then:
\begin{equation}
    \begin{aligned}
        \boldsymbol{z}_h(\boldsymbol{p},X)&=\boldsymbol{z}_h^I(\boldsymbol{p},X)-\{M_{h}^{-1}\}_{N\times |\mathcal{C}_h|}(\{M_{h}^{-1}\}_{|\mathcal{C}_h|\times |\mathcal{C}_h|})^{-1}\{\boldsymbol{z}^I_{h}\}_{|\mathcal{C}_h|\times 1}=(M^{\mathcal C}_h)^{-1}(X\boldsymbol{\beta}_h+\boldsymbol{\xi}-\boldsymbol{p})
    \end{aligned}
\end{equation}

for $M_{\mathcal C_h}^{-1} \equiv M^{-1}_h-\{M^{-1}_{h}\}_{N\times |\mathcal{C}_h|}(\{M_{h}^{-1}\}_{|\mathcal{C}_h|\times |\mathcal{C}_h|})^{-1}\{M^{-1}_{h}\}'_{|\mathcal{C}_h|\times N}$ the effective demand Jacobian associated with set $\mathcal C_h$ for price $\boldsymbol{p}$ and product characteristics $X$. A product whose choice probability has reached zero automatically stops contributing its own- and cross-price responses, while the remaining goods continue to satisfy a linear demand.\\

This gives us an exact decomposition of aggregate demand. Under the same aggregation condition used above for the interior model, we may write aggregate demand as follows:
\begin{equation}\label{eq:aggregate_structural}
\begin{aligned}
\boldsymbol q(\boldsymbol p,X,\boldsymbol{\xi})
&=M^{-1}(X\boldsymbol{\beta}+\boldsymbol{\xi}-\boldsymbol{p})-\sum_h\mu_h \{M^{-1}_{h}\}_{N\times |\mathcal{C}_h|}(\{M_{h}^{-1}\}_{|\mathcal{C}_h|\times |\mathcal{C}_h|})^{-1}\{M^{-1}_h(X\boldsymbol{\beta}_h+\boldsymbol{\xi}-\boldsymbol{p})\}_{|\mathcal{C}_h|\times 1}\\
&=
M^{-1}(X\boldsymbol\beta+\boldsymbol{\xi}-\boldsymbol p)
+
\boldsymbol{\kappa}(\boldsymbol p,\boldsymbol{\xi},X)
\end{aligned}
\end{equation}

for $\boldsymbol{\kappa}(\boldsymbol p,X)$ the aggregate correction generated by consumer-level non-negativity constraints. Unlike unconstrained demand, which only depends on $\mathbb{E}[\boldsymbol{\beta}_h]=\boldsymbol{\beta}$, the corner correction depends on the full distribution of $\boldsymbol{\beta}$ parameters, since heterogeneity in preferences determines which constraints bind. Conditional on a fixed set $\mathcal C_h$, consumer-level demand remains linear, with price derivative $-M_{\mathcal C_h}^{-1}$. Aggregating across consumers, we therefore obtain, wherever the derivatives exist:
\begin{equation}
D_p\boldsymbol q(\boldsymbol p,X,\boldsymbol{\xi})=-\sum_h\mu_h M_{\mathcal C_h}^{-1},\qquad
D_p\boldsymbol\kappa(\boldsymbol p,X,\boldsymbol{\xi})=M^{-1}-\sum_h\mu_h M_{\mathcal C_h}^{-1}
\end{equation}

With finitely many consumers, demand is continuous and piecewise linear, with changes in its Jacobian whenever consumer-level constraints become binding or cease to bind. With continuous heterogeneity, however, these changes need not occur at the same prices across segments. In our context, this gives us smoothness:\\

\textit{\textbf{Proposition 3}: If, at any given prices, there is no positive mass of consumers exactly at the threshold where a product starts or stops being purchased, then aggregate demand $\boldsymbol q(\boldsymbol p,X,\boldsymbol{\xi})$ is continuously differentiable in that region.}\\

Proposition 3 does not impose any particular distribution of consumer preferences or product orderings. However, that there be no positive mass of consumers at thresholds requires threshold-driven changes in the demand Jacobian that are sufficiently dispersed across prices. Such dispersion is especially dependent on heterogeneity in consumer-specific product orderings, distances between alternatives, or characteristic preferences. This is because, from the KKT conditions, the prices at which constraints become binding or cease to bind depend on $\boldsymbol\beta_h$. Continuous heterogeneity in these preferences therefore ensure that no positive mass of consumers changes its binding constraints at exactly the same prices, provided the relevant thresholds vary nontrivially with $\boldsymbol\beta_h$. We explore this next.

\section{Structural Estimation with Market-Level
Data}\label{structural-estimation-with-market-level-data}

The model as developed above requires demand estimation of a piecewise linear system, and the combined estimation of the interior linear demand function with the corner distance densities. In this section, I sketch out one way to identify and estimate each parameter and distribution.\\

Assume we observe only market-level data, e.g. via anonymised transactions over multiple periods: $\boldsymbol q_t$, prices $\boldsymbol p_t$, product characteristics $X$, and excluded instruments $P_t^I$ over periods $t=1,\ldots,T$. For the sake of simplifying the explanation of the estimation procedure, we first set $\boldsymbol\xi_t=\boldsymbol0$, sticking with a pure characteristics model. We end the section with the necessary steps required to generalise $\boldsymbol{\xi}_t$.

\subsection{Parameterising unconstrained linear
demand}\label{parameterising-unconstrained-linear-demand}

One of the first issues to address is the fact that, without consumer-level data, we cannot identify or estimate any consumer-level parameters. We do not observe consumers, their choice sets, or their choice probabilities. Without consumer-level data, the objects associated with those are not parameters which we can separately identify or estimate. Their role is instead to place structure on the aggregate demand function.\\

The first step in estimation is to obtain a parsimonious initial estimate of $M$. To do so, I will rely on the fact that we have allowed $F_h(\boldsymbol{\varepsilon}|X,\boldsymbol{\xi})$, $\mathcal Z_h(\boldsymbol{z}_h,X,\boldsymbol{\xi})$, and ultimately $M_h(X,\boldsymbol{\xi})$, to depend at least partially on $X$. This seems appropriate: more similar goods are presumably more likely to be adjacent in a one-dimensional ordering, and therefore consumers may hold correlated stochastic utility for them.\\

Let $K$ be the number of observed product characteristics and write the QR decomposition:
\begin{equation}
\label{eq:QR_estimation}
X=QR,
\qquad
Q'Q=I_{K}
\end{equation}

where $Q$ is a $N\times K$ orthonormal matrix and $R$ is a $K\times K$ upper triangular matrix. I make use of Lemma 1 from Rodrigues (2026):\\

\textit{\textbf{Lemma 1 (Rodrigues, 2026)}: Any symmetric positive definite matrix $M$ that depends in part on observed product characteristics $X$ and imposes no arbitrary structure on unobserved dimensions admits the following representation:}

\begin{equation}
M(X,\boldsymbol{\xi}_t)=QD(\boldsymbol{\xi}_t)Q'+\rho(\boldsymbol{\xi}_t)(I-QQ')
\end{equation}

\textit{where $D$ is a $K\times K$ symmetric positive-definite matrix and $\rho>0$.}\\

The matrix $D$ summarises how the observed characteristics contribute to the slope of demand and interact with one another. The scalar $\rho$ effectively weighs the relevance of a baseline-differentiation term. This expression is general: $Z$ can be used instead of $X$ as the columns of $Z$ form an orthonormal basis for the column space (or range) of the matrix $X$.\\

Not only is this decomposition of $M$ general, but it is itself microfounded (provided consumers agree on what a product's characteristics are):
\begin{equation}
    \begin{aligned}
        M(X)^{-1}&=\int_{\mathcal{H}}QD_h(\boldsymbol{\xi}_t)^{-1}Q'+\rho_h(\boldsymbol{\xi}_t)^{-1}(I-QQ')\ dh\\
        &=Q[\int_{\mathcal{H}}D_h(\boldsymbol{\xi}_t)^{-1}\ dh]Q'+[\int_{\mathcal{H}}\rho_h(\boldsymbol{\xi}_t)^{-1}\ dh](I-QQ')\\
        &= QD(\boldsymbol{\xi}_t)^{-1}Q'+\rho(\boldsymbol{\xi}_t)^{-1}(I-QQ')
    \end{aligned}
\end{equation}

for $D(\boldsymbol{\xi}_t)=[\int_{\mathcal{H}}D_h(\boldsymbol{\xi}_t)^{-1}\ dh]^{-1}$ and $\rho(\boldsymbol{\xi}_t)=[\int_{\mathcal{H}}\rho_h(\boldsymbol{\xi}_t)^{-1}\ dh]^{-1}$. We collect the free coefficients in $\theta_M$:

\begin{equation}
\boldsymbol\theta_M=\left(\operatorname{vech}(D_1^{-1})',\ldots,
\operatorname{vech}(D_T^{-1})',\rho_1,\ldots,\rho_T\right)'
\end{equation}

After inversion and some re-arranging, $M$ within unconstrained demand may consequently be written as follows:
\begin{equation}
\label{eq:interior_regression}
\begin{aligned}
\boldsymbol m_t(\boldsymbol\beta,\boldsymbol\theta_M)
&\equiv
M^{-1}
(X\boldsymbol\beta-\boldsymbol p_t)=QD_t^{-1}R\boldsymbol\beta
-\frac{1}{\rho_t}\boldsymbol p_t-Q\left(D_t^{-1}-\frac{1}{\rho_t}I\right)
Q'\boldsymbol p_t\\
\end{aligned}
\end{equation}

Implied demand is then:\footnote{The restrictions imposed above on $M$ apply to the candidate parameters used in estimation. A Cholesky decomposition of $D_t$ and a log parameterisation of $\rho_t$ can enforce positivity in the numerical search; the empirical specification may use common $D,\rho$.}

\begin{equation}
\boldsymbol q_t=\boldsymbol m_t(\boldsymbol\beta,\boldsymbol{}\theta_M)
+\boldsymbol\kappa_t(\boldsymbol p_t,X,\boldsymbol\beta,\boldsymbol\theta_M)
+\boldsymbol v_t
\end{equation}

\subsection{Addressing price endogeneity and obtaining starting
values}\label{addressing-price-endogeneity-and-obtaining-starting-values}

For identification, we need to overcome endogeneity which may arise from a correlation between prices and the unobserved error term $\boldsymbol v_t(\boldsymbol p_t,X,\boldsymbol\beta,\boldsymbol\theta_M)$. Let there be a candidate instrument matrix ${P}_{t}^I$, with number of dimensions $J$ equal to the number of unobserved parameters, and define $H_t=\{\mathrm{diag}((\boldsymbol{x}_1)',\ldots,(\boldsymbol{x}_K)',\boldsymbol{p}_{1t}^I)',\ldots,(\boldsymbol{p}_{Jt}^I)'\}\in\mathbb R^{N\times N(K+J)}$.\\

Identification requires sufficient independent moment restrictions and the relevant rank conditions for the chosen parameterisation. We use $\mathbb E[H_t'\boldsymbol v_t]=0$ as our moment conditions. Set the following initial estimate omitting the corner correction. For $\boldsymbol\theta_0=(\boldsymbol\beta',\boldsymbol\theta_M')'$, define $v_t(\boldsymbol\theta_0)$:

\begin{equation}
\boldsymbol v_t(\boldsymbol\theta_0)=\boldsymbol q_t-\boldsymbol m_t(\boldsymbol\theta_0),\qquad
\boldsymbol g(\boldsymbol\theta_0)=\frac1T\sum_{t=1}^T H_t'\boldsymbol v_t(\boldsymbol\theta_0)
\end{equation}

The implied starting values $\boldsymbol\theta_0$ are estimated from the following expression:

\begin{equation}
\widehat{\boldsymbol\theta}_0=\underset{\boldsymbol\theta\in \Theta}{\arg\min}\ 
\boldsymbol g(\boldsymbol\theta)'W_0\boldsymbol g(\boldsymbol \theta)
\end{equation}
where $W_0$ is the starting weighting matrix and $\Theta$ imposes the demand restrictions described above.\\

One concern raised in Gandhi and Nevo (2021) is that it may not be possible to find sufficient (non-collinear) instruments to address endogenous prices when all goods' enter the demand function of all other goods. As can be seen in (\ref{eq:interior_regression}), however, prices enter as a vector; and their direct and indirect impacts - in the second to last and last terms of that equation respectively - can be separated by $QQ'$ in the last term.\\

However, and crucially, when corners are present, the error term includes not just $\boldsymbol{v}_t$ but also $\boldsymbol{\kappa}_t$, and the instruments will therefore not be orthogonal to the omitted correction: $E[H_t'\boldsymbol \kappa_t]\neq\boldsymbol0$. We must characterise the corner correction, doing so only using the market-level data available. The estimate produced in this section supply the starting values for a more general estimator.

\subsection{Accounting for corner solutions}\label{accounting-for-corner-solutions}
If we only observe market-level data, we need to find a way to discipline these corner solutions using only aggregate terms. We start with the following powerful result:\\

\textit{\textbf{Proposition 4:} Any symmetric positive-definite $M^{-1}$ which satisfies $[M^{-1}]_{ij}<0,\ \forall i\neq j$, and $M^{-1}\boldsymbol{1}\geq0$ can be generated by some finite mixture of random-utility-maximising consumers with stochastic utility distribution as specified above.}\\

The trick behind this Proposition is to model the population of consumers using a subset of every conceivable one-dimensional ordering of the alternatives instead of attempting to estimate individual consumer preferences.\\

Here's how we can apply this insight for empirical estimation. Let $\mathcal P$ be the set of one-dimensional orderings of the $N+1$ alternatives, including the outside option. We may use $R$ such simulated permutations, $\pi_1,\ldots,\pi_R$, each representing one possible arrangement of the alternatives along a consumer's ordering. Under uniform draws, the probability that two alternatives are adjacent in any given permutation is:
\begin{equation}\nonumber
    \mathrm{Pr}(i\sim_{\pi_r}j)
    =\frac{2N!}{(N+1)!}
    =\frac{2}{N+1}
\end{equation}

Ideally, we would like to use all permutations in estimation. Where this is not possible, what subset we choose may provide a better or worse fit to the data. To e.g. allow some pairs to be adjacent more frequently than others, we allow in such cases the complete orderings themselves to have different probabilities. These probabilities may be drawn jointly from a symmetric $\mathrm{Dirichlet}(\delta/|\mathcal{P}|,\dots,\delta/|\mathcal{P}|)$, for $\delta$ a parameter for total concentration, which controls how tightly the randomly drawn ordering probabilities cluster around equal probabilities. As $\delta\rightarrow\infty$, the probabilities of the different orderings approach uniformity, and every pair has adjacency probability $2/(N+1)$. As $\delta\rightarrow0$, probability concentrates on a single randomly favoured ordering, making its adjacent pairs much more frequent than other pairs. There are $|\mathcal P|=(N+1)!$ probabilities, each with Dirichlet parameter $\delta/|\mathcal P|$, and they sum to one. We draw this probability distribution once, then draw $\pi_1,\ldots,\pi_R$ independently conditional on it.\footnote{The permutations can be generated without enumerating $\mathcal P$. Draw $\pi_1$ uniformly. At draw $r\geq2$, draw a uniform permutation with probability $\delta/(\delta+r-1)$; otherwise, select one of the previous $r-1$ draws uniformly and repeat its ordering. Selection is among previous draws, including repetitions. This generates the same joint distribution as the Dirichlet specification (Blackwell and MacQueen, 1973).} Retain the draws throughout estimation except when estimating $\delta$ directly.\\

Let $\hat n_{ij}$ be the number of distinct permutations in which $i,j$ are adjacent:
\begin{equation}
\label{eq:adjacency_frequency}
    \hat n_{ij}
    =
    \sum_{r=1}^{R}
    1\{i\sim_{\pi_r(\delta)}j\}
\end{equation}

where $1\{i\sim_{\pi_r(\delta)}j\}$ indicates whether $i$ and $j$ are adjacent in permutation $\pi_r$. This definition applies to all pairs, including those involving the outside option.\footnote{Note that the construction below requires $\hat n_{ij}>0$ whenever $w_{ij,t}>0$.} For every adjacent pair under ordering $\pi_r$, define:
\begin{equation}
\label{eq:simulated_distance}
    d_{r,ij}=\lambda_{\pi_r}\frac{\hat n_{ij}}{w_{ij,t}}
\end{equation}
for $\lambda_{\pi_r}>0$ a permutation weight satisfying $\sum_{r=1}^R \lambda_{\pi_r}=1$. This implies the following response matrix construction, from ($\ref{responsematrixx}$):
\begin{equation}
\label{eq:simulated_L}
\begin{aligned}
        L_{r,t}&=\sum_{0\leq i<j\leq N,\ i\sim_{\pi_r(\delta)}j}\frac{1}{d_{r,ij}}(\boldsymbol e_i-\boldsymbol e_j)    (\boldsymbol e_i-\boldsymbol e_j)'\\
        &=\sum_{0\leq i<j\leq N,\ i\sim_{\pi_r(\delta)}j}\frac{w_{ij,t}}{\lambda_{\pi_r}\hat n_{ij}}(\boldsymbol e_i-\boldsymbol e_j)    (\boldsymbol e_i-\boldsymbol e_j)'\\
\end{aligned}
\end{equation}

The inside-good block of $L_{r,t}$ defines the response matrix $M_{r}^{-1}$ associated with simulated ordering $\pi_r$. This construction has the following exact finite-simulation property:
\begin{equation}\label{ellell}
\begin{aligned}
    \sum_{r=1}^{R}\lambda_{\pi_r}L_{r,t}
    &=\sum_{r=1}^R\sum_{i\sim_{\pi_r(\delta)}j}\frac{w_{ij,t}}{\hat n_{ij}}(\boldsymbol e_i-\boldsymbol e_j)    (\boldsymbol e_i-\boldsymbol e_j)'\\
    &=
    \sum_{0\leq i<j\leq N}
    \frac{w_{ij,t}}{\hat n_{ij}}
    \left[
        \sum_{r=1}^{R}
        1\{i\sim_{\pi_r}j\}
    \right]
    (\boldsymbol e_i-\boldsymbol e_j)
    (\boldsymbol e_i-\boldsymbol e_j)'\\
    &=
    \sum_{0\leq i<j\leq N}
    w_{ij,t}
    (\boldsymbol e_i-\boldsymbol e_j)
    (\boldsymbol e_i-\boldsymbol e_j)'\\
    &=L_t
\end{aligned}
\end{equation}

The manner in which we weigh each permutation cancels out - our results generalise across any such weight. Taking the inside-good block therefore gives:
\begin{equation}
\label{eq:exact_simulated_M}
    \sum_{r=1}^{R}\lambda_{\pi_r}M_{r}^{-1}
    =M^{-1}
\end{equation}

Thus, the simulated response matrices reproduce the candidate aggregate matrix exactly, even for finite $R$. It
supplies the $M_{r,t}^{-1}$ used to calculate consumer-level corner solutions.\\

% In summary, these permutations can provide a structured decomposition of the aggregate unconstrained demand from which to compute the consumer-level corner corrections. $L_t$ determines the aggregate substitution weights $w_{ij,t}$, while demand across prices determines how much each ordering contributes to these weights. For an ordering $\pi$ and a pair $i,j$ adjacent in that ordering, this contribution is $\sum_{r}\frac{\lambda_{\pi_r}}{d_{r,ij}}$. This ratio, summed across repetitions of the same ordering, is then determined by demand across the stated price range.\\

While the permutation method above addresses heterogeneity in the consumer-level $M^{-1}_h$ terms, the location of a consumer's choice probabilities relative to corner boundaries depends also on preferences parameters $\boldsymbol{\beta}_h$. Our corner correction depends not just on $\boldsymbol{\beta}$, but its distribution in the consumer population. Notably, both the distributions of preference heterogeneity and permutation need to be specified if we wish to obtain uniqueness of the simulated corner correction term.\\

% We can now specify where latent demand lies relative to the corner boundaries. For this purpose define aggregate zero-price interior demand across goods:
% \begin{equation}
% \label{eq:sbar}
% \overline{\boldsymbol s}_t
% \equiv
% M^{-1}X\boldsymbol\beta
% =
% QD^{-1}R\boldsymbol\beta.
% \end{equation}

% The ordering construction determines heterogeneity in the consumer-level initial demand Jacobian $M_h^{-1}$, but the location of a segment relative to its corner boundaries also depends on its characteristic preferences $\boldsymbol\beta_h$. Interior aggregate demand depends only on the conditional mean of these preferences. Corner solutions, by contrast, depend on their distribution around that mean. We therefore introduce a random-coefficient specification.\\

For each simulated ordering $\pi_r$, define $\boldsymbol\beta_{r\ell}$ as follows:
\begin{equation}
\label{eq:random_coefficients}
\boldsymbol\beta_{r\ell}=\boldsymbol\beta+\Gamma\boldsymbol\zeta_{r\ell}\qquad 
E[\boldsymbol\zeta_{r\ell}]=\boldsymbol0
\qquad 
E[\boldsymbol\zeta_{r\ell}\boldsymbol\zeta_{r\ell}']=I_{K}
\end{equation}

the $\Gamma$ a $K\times K$ the Cholesky factor of the variance-covariance matrix of $\boldsymbol{\beta}_{r\ell}$. The mean coefficient remains $\boldsymbol\beta$, while $\Gamma$ governs the heterogeneity in characteristic tastes which is relevant for the corner correction. For simplicity, let $\Gamma=\operatorname{diag}(\gamma_1,\ldots,\gamma_{K})$, for $\gamma_k\geq0,\ \forall k$ as a baseline. Let $\ell=1,\ldots,S$ index $\boldsymbol{\zeta}_{r\ell}$ draws from a conditional distribution - I use $\boldsymbol{\zeta}_{r\ell,k}\sim U[-\sqrt{3},\sqrt{3}]$, constructed such that $(1/S)\sum_\ell\boldsymbol{\zeta}_{r\ell}=0$.\footnote{A bounded symmetric mixing distribution provides a convenient specification of heterogeneity in preferences for characteristics, with boundedness in particular making the assumption of slackness in the outside option condition possible. For example, the components of $\boldsymbol\zeta_{r\ell}$ may be independently distributed, e.g. $\zeta_{r,\ell, k}\sim U[-\sqrt{3},\sqrt{3}]$ which gives zero mean and unit variance. The simulation may be constructed so that the conditional-mean restriction is satisfied exactly for finite $S$, for example by conditioning the draws on $\frac{1}{S}\sum_{\ell=1}^{S}\boldsymbol\zeta_{r\ell}=\boldsymbol0$ for every $r$.} Retain the underlying draws throughout. It follows from (\ref{eq:random_coefficients}) that $\frac{1}{S}\sum_{\ell=1}^{S}\boldsymbol\beta_{r\ell}=\boldsymbol\beta$.\\

At observed prices, averaging over unconstrained demands for each  simulated ordering and random coefficient draws $(r,\ell,t)$, is therefore:

\begin{equation}\label{eq:simulated_interior_demand}
\begin{aligned}
\frac{1}{S}\sum_{r=1}^{R}\lambda_{\pi_r}\sum_{\ell=1}^{S}\boldsymbol z_{r,\ell,t}^{I}&=\frac{1}{S}\sum_{r=1}^{R}\lambda_{\pi_r}M_{r,t}^{-1}\left(X\boldsymbol\beta_{r\ell}-\boldsymbol p_t\right)\\
&=M^{-1}\left(X\boldsymbol\beta-\boldsymbol p_t\right)\\
\end{aligned}
\end{equation}

In other words, the random-coefficient simulation reproduces exactly the aggregate unconstrained demand system implied by the initial model, while allowing the latent demands around that mean to be heterogeneous. This is key for our following step: completing the argument that we can use aggregate terms to fully constraint the corner correction, and therefore eliminate price endogeneity. It is key because unconstrained demand enters the corner correction directly, as we saw in e.g. (\ref{eq:aggregate_structural}).\\

We can exploit the one-dimensional structure of the idiosyncratic valuations underlying each simulated permutation to obtain the constrained choice probabilities directly, without iterating over candidate corner sets.\\

The positions $\alpha_{ir}<0$ for all inside goods and $\alpha_{0,r,t}=0$ can be recovered from the distances $d_{r,ij}$ defined in (\ref{eq:simulated_distance}), up to the normalisation of the outside option. For two alternatives $i$ and $i+1$, with $\alpha_{ir}<\alpha_{i+1,r}$, their utility functions intersect at:
\begin{equation}
\epsilon_{i,i+1,r,\ell,t}^{*}=\frac{\nu_{i,r,\ell,t}-\nu_{i+1,r,\ell,t}
}{\alpha_{i+1,r}-\alpha_{ir}}=\frac{\nu_{i,r,\ell,t}-\nu_{i+1,r,\ell,t}}{d_{r,i,i+1}}
\label{eq:simulated_utility_intersection}
\end{equation}

To obtain the corner corrections, start from the alternative with lowest $\alpha$ in $\pi_r$, and compare each successive alternative. If their intersection occurs before the point at which the first alternative becomes optimal - e.g. if $\epsilon^*_{i,i+1}\leq\epsilon^*_{i-1,i}$ - that alternative $i$ is removed, and the comparison is repeated with its predecessor. Otherwise, the new alternative is added to an \textit{envelope}.\footnote{By envelope here we mean the upper envelope of the lines that can be seen at the top of the left and right sub-graphs of Figure 1, which correspond to what a given consumer purchases at each $\varepsilon$.}\\

Let $\mathcal B_{r,\ell,t}=(b_1,\ldots,b_m)$ be this envelope, the resulting sequence of retained alternatives, ordered by increasing $\alpha_{ir}$. Define the intersection between consecutive retained alternatives as:
\begin{equation}
\tau_{k,r,\ell,t}=\frac{\nu_{b_k,r,\ell,t}-\nu_{b_{k+1},r,\ell,t}}{\alpha_{b_{k+1},r}-\alpha_{b_k,r}},\qquad k=1,\ldots,m-1
\end{equation}

with $\tau_{0,r,\ell,t}=-\infty$ and $\tau_{m,r,\ell,t}=+\infty$. To derive the corresponding choice probabilities, note that, by construction of the envelope, each retained alternative $b_k$ maximises utility between its intersections with the preceding and succeeding retained alternatives. E.g., since $\alpha_{b_{k-1},r}<\alpha_{b_k,r}<\alpha_{b_{k+1},r}$, we have:
\begin{equation}\nonumber
\begin{aligned}
u_{b_k,r,\ell,t}(\epsilon)&\geq u_{b_{k-1},r,\ell,t}(\epsilon)\quad\Leftrightarrow\quad\epsilon\geq\tau_{k-1,r,\ell,t}\\
u_{b_k,r,\ell,t}(\epsilon)&\geq u_{b_{k+1},r,\ell,t}(\epsilon)\quad\Leftrightarrow\quad\epsilon\leq\tau_{k,r,\ell,t}
\end{aligned}
\end{equation}
with the first and last alternatives bounded by the uniform distribution. The set of idiossyncratic errors for which a given alternative $b_k$ maximises utility over all available alternatives is:
\begin{equation}
\begin{aligned}
\mathcal E_{k,r,\ell,t}
&=\left\{\epsilon\in[0,1]:
\tau_{k-1,r,\ell,t}\leq\epsilon
\leq\tau_{k,r,\ell,t}\right\}\\
&=[0,1]\cap
[\tau_{k-1,r,\ell,t},\tau_{k,r,\ell,t}]\\
&=\left[
\max\{0,\tau_{k-1,r,\ell,t}\},
\min\{1,\tau_{k,r,\ell,t}\}
\right]
\end{aligned}
\end{equation}

The corresponding choice probability is simply the length of this interval:
\begin{equation}
\begin{aligned}
z_{b_k,r,\ell,t}
&=\Pr\left(\epsilon\in\mathcal E_{k,r,\ell,t}\right)\\
&=\int_0^1
\boldsymbol{1}\left\{
\tau_{k-1,r,\ell,t}\leq\epsilon
\leq\tau_{k,r,\ell,t}
\right\}d\epsilon\\
&=\max\left\{
\min\{1,\tau_{k,r,\ell,t}\}
-\max\{0,\tau_{k-1,r,\ell,t}\},0
\right\}
\end{aligned}
\end{equation}

The clipping at zero and one reflects the bounds of the support of $\epsilon$. Alternatives which never maximise utility over this interval receive zero demand. The realised corner set is therefore obtained immediately as
$\mathcal C_{r,\ell,t}= \{i\in\mathcal N\setminus\{0\}:z_{i,r,\ell,t}=0\}$. The exact permutation-draw-level corner correction follows directly from the difference between constrained and unconstrained demand $\boldsymbol\kappa_{r,\ell,t}=\boldsymbol z_{r,\ell,t}-\boldsymbol z_{r,\ell,t}^{I}$.\\

The corresponding predicted aggregate demand and simulated corner correction are thus:
\begin{equation}
\label{eq:simulated_K}
\widehat{\boldsymbol q}_{R,S,t}(\boldsymbol\beta,\boldsymbol\theta_M,\boldsymbol\gamma,\delta)
=\boldsymbol m_t(\boldsymbol\beta,\boldsymbol\theta_M)
+\widehat{\boldsymbol\kappa}_{R,S,t}(\boldsymbol\beta,\boldsymbol\theta_M,\boldsymbol\gamma,\delta)
\qquad
\widehat{\boldsymbol{\kappa}}_{R,S,t}=\frac{1}{S}\sum_{r=1}^{R}\lambda_{\pi_r}\sum_{\ell=1}^{S}\boldsymbol \kappa_{r,\ell,t}
\end{equation}

Note that the second equation is independent of $\lambda_{r\ell}$, cancelling out in the same way as they did above due to $\kappa_{r,\ell,t}\propto\frac{1}{\lambda_{\pi_r}}$.\footnote{This implies that once the set of permutations is fixed, changing their probabilities has no effect on aggregate demand or the corner correction. In other words, if we could calculate every sample permutation, $\delta$ would no longer be a source of variation and could be set to $\delta=1$. Instead, as mentioned above, we keep $\delta$ and instead simulate only a sample of all possible permutations for computational simplicity.}\\

\subsection{Estimation procedure}\label{estimation-procedure}

Organise the parameters as follows: $\boldsymbol\theta_1=\boldsymbol\beta,\ 
\boldsymbol\theta_2=(\boldsymbol\theta_M',\boldsymbol\gamma',\delta)'.$ The $\boldsymbol{\theta}_2$ parameters are estimated in the outer loop. At each candidate $\boldsymbol\theta_2$, the inner loop estimates $\boldsymbol\beta$ by GMM. Define the simulated residual and corresponding moments, respectively:

\begin{equation}
\begin{aligned}
\boldsymbol v_{R,S,t}(\boldsymbol\beta,\boldsymbol\theta_2)
&=\boldsymbol q_t-\widehat{\boldsymbol q}_{R,S,t}(\boldsymbol\beta,\boldsymbol\theta_2)\\
\boldsymbol g_{R,S,T}(\boldsymbol\beta,\boldsymbol\theta_2)
&=\frac1T\sum_{t=1}^T H_t'\boldsymbol v_{R,S,t}(\boldsymbol\beta,\boldsymbol\theta_2)\\
\end{aligned}
\end{equation}

For a fixed weighting matrix, the inner estimate is

\begin{equation}
\widehat{\boldsymbol\beta}(\boldsymbol\theta_2)=\underset{\boldsymbol\beta}{\arg\min}
\ \boldsymbol g_{R,S,T}(\boldsymbol\beta,\boldsymbol\theta_2)'W_T
\boldsymbol g_{R,S,T}(\boldsymbol\beta,\boldsymbol\theta_2)
\end{equation}

Each trial $\boldsymbol\beta$ requires calculating $\boldsymbol\beta_{r\ell}$, $\boldsymbol\nu_{r\ell,t}=X\boldsymbol{\beta}_{r\ell}-\boldsymbol{p}_t$, $\boldsymbol z^I_{r\ell,t}$, $\boldsymbol z_{r\ell,t}$, and the resulting moments. The inner GMM problem is generally nonlinear because $\boldsymbol\beta$ changes the corner sets. \\

The outer estimate is:
\begin{equation}
\widehat{\boldsymbol\theta}_2=\underset{\boldsymbol\theta_2\in\Theta}{\arg\min}
\ \boldsymbol g_{R,S,T}(\widehat{\boldsymbol\beta}(\boldsymbol \theta_2),\boldsymbol\theta_2)'W_T
\boldsymbol g_{R,S,T}(\widehat{\boldsymbol\beta}(\boldsymbol\theta_2),\boldsymbol\theta_2)
\end{equation}

An evaluation of this outer objective proceeds as follows:

\begin{enumerate}
    \item Fix a candidate $\boldsymbol\theta_2$, including $D_t,\rho_t,\boldsymbol\gamma,\delta$
    \item Construct $M_t^{-1}$ and $M_{r,t}^{-1}$, and draw and form $\Gamma\boldsymbol{\zeta}_{r,\ell,t}$
    \item Estimate $\widehat{\boldsymbol\beta}(\boldsymbol\theta_2)$ by GMM, updating $\boldsymbol z_{r\ell,t}$ and $\widehat{\boldsymbol\kappa}_{R,S,t}$ at each trial $\boldsymbol\beta$
    \item Return the resulting $\boldsymbol v_{R,S,t}$, $\boldsymbol g_{R,S,T}$, and minimised criterion to the outer loop
\end{enumerate}

The outer loop updates $\boldsymbol\theta_2$ and repeats these steps. The initial estimate from Section \label{endogeneity} supplies starting values. At convergence, estimate the covariance of the moment contributions $H_t'v_{R,S,t}$, allowing for the dependence in the observations. Update $W_T$ and repeat both loops to obtain the conventional two-step GMM.

\subsection{Including income effects}\label{incincome}
The structural estimation above can be extended to incorporate the income effects developed earlier with only a minor modification. Recall that, after the normalisation introduced there, consumer-level unconstrained demand can be written as:

\begin{equation}
    \boldsymbol{z}_h^I=M_h^{-1}(X\boldsymbol{\beta}_h+\eta_h\mathbf 1-\boldsymbol{p})
\end{equation}

where $\eta_h$ captures heterogeneity across consumers in the utility contribution associated with residual income. Since neither consumers nor individual incomes are observed in the market-level data, we instead directly specify a joint distribution for the consumer-level characteristic preferences and income-related utility term. Generalising the random-coefficient specification above, we get:

\begin{equation}
\begin{pmatrix}
\boldsymbol{\beta}_h\\
\eta_h
\end{pmatrix}
=
\begin{pmatrix}
\boldsymbol{\beta}\\
\eta
\end{pmatrix}
+
\Gamma\boldsymbol{\zeta}_h
\end{equation}

where $\Gamma$, now a $(K+1)\times(K+1)$ matrix, retains its earlier interpretation. Assume the $\boldsymbol{\zeta}$ distribution has the following moments:

\begin{equation}
E[\boldsymbol{\zeta}_h]=0\qquad
E[\boldsymbol{\zeta}_h\boldsymbol{\zeta}_h']=I_{K+1}\qquad
\operatorname{Var}
\begin{pmatrix}
\boldsymbol{\beta}_h\\
\eta_h
\end{pmatrix}
=
\Gamma\Gamma'
\equiv
\Sigma
=
\begin{pmatrix}
\Sigma_{\beta\beta} & \Sigma_{\beta\eta}\\
\Sigma_{\eta\beta} & \gamma_\eta^2
\end{pmatrix}
\end{equation}

Estimating the joint distribution directly is advantageous: the vector $\Sigma_{\beta\eta}$ allows the income-related utility term to be correlated with each element of $\boldsymbol{\beta}_h$. Thus, for example, consumers with systematically larger $\eta_h$ may also place systematically greater weight on particular observed product characteristics. The remaining elements of $\Sigma_{\beta\beta}$ continue to allow arbitrary taste heterogeneity unrelated to $\eta_h$.\\

Implementation then proceeds exactly as above. For each simulated ordering $r$ and heterogeneity draw $\ell$, jointly draw:
\begin{equation}
\begin{pmatrix}
\boldsymbol{\beta}{r\ell}\\
\eta_{r\ell}
\end{pmatrix}
=
\begin{pmatrix}
\boldsymbol{\beta}\\
\eta
\end{pmatrix}
+
\Gamma\boldsymbol{\zeta}_{r\ell}
\end{equation}

The same procedure can then be applied to $z^I_{r,\ell,t}$ to determine the set of binding non-negativity constraints and the associated corner correction. No other modification to the simulated corner estimator is required. Once estimated, the joint distribution of $\{\beta,\eta\}$ can be held fixed when evaluating counterfactual changes in prices or product characteristics, with the corresponding corner sets recomputed under each counterfactual.\\

How we approach income effects has implications for the model's status as an incomplete demand system. Over time, $g(\cdot)$ is likely to vary as omitted goods vary in e.g. prices. This variation can be incorporated into the model by allowing $\eta$ to be time-varying through $a^1$ and $a^0$, which preserves time invariance in the demand Jacobian and characteristics parameters. Furthermore, note that variation in the prices or characteristics of omitted products may enter the regression error when those products substitute with inside goods. If we wish to keep an incomplete demand system, the instrument orthogonality condition requires care to ensure that instruments for inside-product prices are not correlated with such omitted variation.

\subsection{Computational complexity}\label{computational-complexity}

Conditional on construction of the candidate matrices and permutations, calculating $X\boldsymbol\beta_{r\ell}$ costs $O(NK)$ for each simulated consumer. The envelope algorithm costs $O(N)$ because each alternative enters and leaves the envelope at most once. Multiplication by $M_{r,t}^{-1}$ also costs $O(N)$. Thus, a conservative bound for one evaluation of the baseline simulated demand system is $O\left(TRSN(K+1)\right)$.\\

At a fixed outer candidate, terms such as $X\Gamma\boldsymbol\zeta_{r\ell}$ can be retained while the inner loop changes $\boldsymbol\beta$. This reduces the number of repeated evaluations. The total estimation cost also depends on the numbers of inner and outer evaluations and on preparation of the matrices. The computations for different $(t,r,\ell)$ can be carried out in parallel conditional on the candidate parameters.\\

For comparison, let $I$ be the average number of iterations required for the BLP inversion to match predicted market shares to their observed counterparts. Holding constant the total number of simulated market-consumer evaluations $C$, a standard simulated BLP implementation requires $O(CNK)$ operations to construct the individual utility indices and $O(CNI)$ operations to perform the share inversion. Its corresponding computational complexity is therefore $O(CN(K+I))$, where the full model presented in this paper is correspondingly $O(CN(K+1))$.\\

The distinction is not merely the number of operations, but also their application. In the full model, the $TRS$ simulated corner problems are independent conditional on the demand parameters, and can therefore be evaluated in parallel before their corrections are aggregated. BLP can similarly parallelise calculations across simulated consumers and markets, but successive share-inversion iterations within each market depend on the preceding iteration. The inversion step furthermore introduces simulation error to BLP estimation (Berry, Linton, and Pakes, 2004), requiring additional work to address.\\

In some settings, this confers it some computational advantages over alternative demand models such as BLP. As such, it may be of interest particularly in applications where computational complexity is a major bottleneck. We discuss examples in the next section.

\subsection{Allowing for unobserved product characteristics}\label{extension-unobserved-product-characteristics}

To complete this section, we now re-introduce $\boldsymbol\xi_t$, common across consumers within market $t$, such that $\boldsymbol\nu_{ht}=X\boldsymbol\beta_h+\boldsymbol\xi_t-\boldsymbol p_t$. This term changes both the unconstrained probabilities and the corner sets. We will account for it by specifying its conditional distribution and averaging predicted demand over simulated draws. Unlike in BLP, $\boldsymbol{\xi}_t$ is not recovered directly.\\

We start by first fitting a price equation, e.g. via polynomial series expansion:
\begin{equation}
    \boldsymbol{p}_t=w(X,\boldsymbol{P}_t^I, D_t, \rho_t)+\tilde{\boldsymbol{p}}_t
\end{equation}
followed by a functional form and distribution assumption:
\begin{equation}
    \boldsymbol{\xi}_t=\lambda\hat{\tilde{\boldsymbol{p}}}_t+\sigma\tilde{\boldsymbol{\xi}}_t
\end{equation}
for $\tilde{\boldsymbol{\xi}}_t\mid X,\boldsymbol P_t^I,D_t,\rho_t,
\tilde{\boldsymbol p}_t\sim N(\tilde{\boldsymbol{\xi}},1)$. While more restrictive than BLP, assuming so is not new (see e.g., Jiang, Manchanda, and Rossi, 2009). The coefficients of $\pi$ are estimated when the candidate $D_t,\rho_t$ change. We assume that $\hat{\tilde{\boldsymbol{p}}}_t$ captures the additional dependence of the conditional distribution of $\boldsymbol\xi_t$ on $D_t,\rho_t$.\\

For identification, I set $\sigma>0$ and propose the separation of $\boldsymbol{\beta}$ from $\boldsymbol{\tilde\xi}$ by imposing an identifying restriction directly: $Q'\tilde{\boldsymbol\xi}=\boldsymbol0$. The restriction leaves $N-K$ degrees of freedom components, in addition to the $K$ coefficients in $\boldsymbol\beta$. Simulate draws $\boldsymbol\xi_{b,t}$, $b=1,\ldots,B$, once. At each iteration in the inner loop, calculate:

\begin{equation}
\boldsymbol\xi_{b,t}
=\lambda\widehat{\tilde{\boldsymbol p}}_t
+\sigma\tilde{\boldsymbol\xi}_{b,t}
\qquad
\boldsymbol\nu_{r\ell b,t}
=X(\boldsymbol\beta+\Gamma\boldsymbol\zeta_{r\ell})
+\boldsymbol\xi_{b,t}-\boldsymbol p_t
\end{equation}

Each draw $\boldsymbol\xi_{b,t}$ is common to all $(r,\ell)$. Use $\boldsymbol\nu_{r\ell b,t}$ both to calculate $\boldsymbol z^{I}_{r\ell b,t}=M_{r,t}^{-1}\boldsymbol\nu_{r\ell b,t}$ and to obtain $\boldsymbol z_{r\ell b,t}$ from the envelope algorithm. The correction for that draw is $\boldsymbol\kappa_{r\ell b,t}=\boldsymbol z_{r\ell b,t}-\boldsymbol z^{I}_{r\ell b,t}$. Average predicted demand over the draws is then:
\begin{equation}
\widehat{\overline{\boldsymbol q}}_{B,R,S,t}(\boldsymbol{\vartheta}_1,\boldsymbol{\vartheta}_2)
=\frac1{BS}\sum_{b=1}^B\sum_{r=1}^R\lambda_{\pi_r}
\sum_{\ell=1}^S\boldsymbol z_{r\ell b,t}
\end{equation}

for $\boldsymbol{\vartheta}_1=(\boldsymbol\beta',\tilde{\boldsymbol\xi}')'$ and $\boldsymbol{\vartheta}_2=(\boldsymbol\theta_M',\boldsymbol\gamma',\delta,\lambda,\sigma)'$. Much like before, we use the following simulated residuals and moments:
\begin{equation}
\begin{aligned}
\boldsymbol v_{B,R,S,t}(\boldsymbol{\vartheta}_1,\boldsymbol{\vartheta}_2)
&=\boldsymbol q_t-
\widehat{\overline{\boldsymbol q}}_{B,R,S,t}(\boldsymbol{\vartheta}_1,\boldsymbol{\vartheta}_2)\\
\boldsymbol g^\xi_T(\boldsymbol{\vartheta}_1,\boldsymbol{\vartheta}_2)
&=\frac1T\sum_{t=1}^T H_t'\boldsymbol v_{B,R,S,t}(\boldsymbol{\vartheta}_1,\boldsymbol{\vartheta}_2)\\
\end{aligned}
\end{equation}

The inner loop estimates $\boldsymbol\vartheta_1$ and the outer loop estimates $D_t,\rho_t,\boldsymbol\gamma,\delta,\lambda,\sigma$, respectively:

\begin{equation}
\begin{aligned}
    \widehat{\boldsymbol{\vartheta}}_1(\boldsymbol{\vartheta}_2)
=\underset{Q'\tilde{\boldsymbol\xi}=\boldsymbol0}{\arg\min}\ 
g^\xi_T(\boldsymbol{\vartheta}_1,\boldsymbol{\vartheta}_2)'W_Tg^\xi_T(\boldsymbol{\vartheta}_1,\boldsymbol{\vartheta}_2)\\
\widehat{\boldsymbol{\vartheta}}_2
=\underset{\boldsymbol{\vartheta}_2\in\varTheta}{\arg\min}\ 
g^\xi_T\left(\widehat{\boldsymbol{\vartheta}}_1(\boldsymbol{\vartheta}_2),\boldsymbol{\vartheta}_2\right)'W_Tg^\xi_T\left(\widehat{\boldsymbol{\vartheta}}_1(\boldsymbol{\vartheta}_2),\boldsymbol{\vartheta}_2\right)
\end{aligned}
\end{equation}
and we can proceed as above.\\

\section{Applications}\label{applications}
\subsection{Welfare analysis}\label{welfare-analysis}

We can use the demand system estimated in the previous section to evaluate counterfactual changes in consumer welfare. We consider a setting without income effects once again, though doing so requires a minimal extension in the same line as the previous section. Crucially, I will show how welfare analysis does not require us to interpret the simulated consumer draws used in estimation as observed or separately identified objects. We observe only market-level prices, characteristics, and quantities, and care only about aggregate consumer welfare.\\

From before:
\begin{equation}
\boldsymbol q(\boldsymbol p,X)=M^{-1}(X)(X\boldsymbol \beta-\boldsymbol p)+\boldsymbol \kappa(\boldsymbol p,X)
\end{equation}

where $\boldsymbol \kappa(\boldsymbol p,X)$ is the aggregate correction generated by consumer-level corner solutions. The discrete-choice microfoundation gives this demand system a corresponding surplus representation. For a simulated ordering $r$ and preference draw $\ell$, let:
\begin{equation}
\boldsymbol v_{r\ell}(\boldsymbol p,X)=X\boldsymbol \beta_{r\ell}-\boldsymbol p\qquad M_r(X)=\left[M_r^{-1}(X)\right]^{-1}
\end{equation}

Note that it follows from (\ref{ellell}) and the later derivations in Section \ref{accounting-for-corner-solutions} that $M_r^{-1}$ is a function of $\lambda_{\pi_r}$, which we can keep track of by writing $M_r=\lambda_{\pi_r}\widetilde M_r$.\\

The inversion required to obtain $M_r$ is made computationally parsimonious through the low-dimensional approach using $X$ expounded in Lemma 1. The expected surplus associated with this draw is:
\begin{equation}
V_{r\ell}(\boldsymbol p,X)=\max_{\boldsymbol z\geq \boldsymbol 0,\ \boldsymbol 1'\boldsymbol z\leq 1}\left\{\boldsymbol v_{r\ell}(\boldsymbol p,X)'\boldsymbol z-\frac{\lambda_{\pi_r}}{2}\boldsymbol z'\widetilde M_r(X)\boldsymbol z\right\}
\end{equation}

The maximiser is exactly the choice-probability vector $\boldsymbol z_{r\ell}(\boldsymbol p,X)$ obtained from the Karush-Kuhn-Tucker problem used to construct the corner correction in the previous section. Under these conditions, note that $V_{r\ell}(\boldsymbol p,X)=\frac{1}{\lambda_{\pi_r}}\widetilde V_{r\ell}(\boldsymbol p,X)$, for $\widetilde V_{r\ell}(\boldsymbol p,X)$ independent of the permutation weight.\\

Aggregate consumer surplus is then approximately:
\begin{equation}
W(\boldsymbol p,X)=\frac{1}{S}\sum_{r=1}^R\lambda_{\pi_r}\sum_{\ell=1}^SV_{r\ell}(\boldsymbol p,X)=\frac{1}{S}\sum_{r=1}^R\sum_{\ell=1}^S\widetilde V_{r\ell}(\boldsymbol p,X)
\end{equation}
which confirms the independence from permutation weights as above.\\

For the sake of example, consider a price counterfactual. By the envelope theorem:
\begin{equation}
\nabla_{\boldsymbol p} \widetilde V_{r\ell}(\boldsymbol p,X)=-\boldsymbol  z_{r\ell}(\boldsymbol p,X)\quad\Rightarrow\quad
\nabla_{\boldsymbol p} W(\boldsymbol p,X)=-\boldsymbol q(\boldsymbol p,X)
\end{equation}

Consequently, for a change from $\boldsymbol p^0$ to $\boldsymbol p^1$, holding $X$ fixed, we obtain:
\begin{equation}
\Delta W_p=-\int_0^1\boldsymbol q\left(\boldsymbol p^0+ s\Delta \boldsymbol p,X\right)'\Delta \boldsymbol p\ ds
\qquad \Delta \boldsymbol p\equiv \boldsymbol p^1-\boldsymbol p^0
\end{equation}

Notably, this is independent of anything consumer-specific. The discrete-choice microfoundation guarantees that this line integral is a difference in surplus rather than a path-dependent area under an arbitrary system of demands. Substituting the estimated demand decomposition yields:
\begin{equation}
\Delta W_p=-\left[M^{-1}(X)(X\boldsymbol \beta-\boldsymbol p^0)\right]'\Delta \boldsymbol p+\frac{1}{2}\Delta \boldsymbol p'M^{-1}(X)\Delta \boldsymbol p-\int_0^1\boldsymbol \kappa(\boldsymbol p^0+s\Delta \boldsymbol p,X)'\Delta \boldsymbol p\ ds
\end{equation}

The estimated structural model supplies $\boldsymbol{\kappa}(\boldsymbol{p},X)$ at each point along the path, so the remaining integral can be evaluated numerically.\\

Let us consider another counterfactual, changes in product characteristics. This requires slightly more structure; in general, it cannot be welfare-valued from aggregate quantities alone. Holding $M_r$ fixed, a characteristic change has the draw-level envelope effect:
\begin{equation}
dV_{r\ell}
=
\boldsymbol z_{r\ell}'\,dX\,\boldsymbol \beta_{r\ell}.
\end{equation}

With heterogeneous $\boldsymbol \beta_{r\ell}$:
\begin{equation}
    \frac{1}{S}\sum_{r=1}^R\lambda_{\pi_r}\sum_{\ell=1}^S\boldsymbol z_{r\ell}'dX\boldsymbol \beta_{r\ell}\neq q'dX\boldsymbol \beta
\end{equation}

Aggregate demand reveals how many consumers choose each product, but not which characteristic coefficients are attached to those consumers. The substitution matrix itself changes with characteristics, as permitted by $M_r(X)$.\\

Thankfully, this is precisely where the structural estimation in the preceding section becomes useful. For a general counterfactual from $(\boldsymbol p^0,X^0)$ to $(\boldsymbol p^1,X^1)$, we hold fixed the underlying simulation draws $(r,\ell)$, including their preference coefficients, but recompute $M_r(X)$ and the associated corner set. Lastly, we calculate:
\begin{equation}
\Delta W=\frac{1}{S}\sum_{r=1}^R\lambda_{\pi_r}\sum_{\ell=1}^S\left[V_{r\ell}(\boldsymbol p^1,X^1)-V_{r\ell}(\boldsymbol p^0,X^0)\right]=\frac{1}{S}\sum_{r=1}^R\sum_{\ell=1}^S\left[\widetilde V_{r\ell}(\boldsymbol p^1,X^1)-\widetilde V_{r\ell}(\boldsymbol p^0,X^0)\right]
\end{equation}

This gives the model-implied welfare effect of both price and characteristic changes while preserving the heterogeneity that was used to fit aggregate corner behaviour.\\

In conclusion, our estimated multi-product aggregate demand system can be used for welfare analysis regardless of whether the relevant counterfactual remains in the interior or crosses onto a case where one or more goods receive zero demand.

\subsection{Multi-product pricing and cost pass-through}\label{pricingpassthrough}

The model above can be combined with a model of firm behaviour to deliver optimal prices, markups, and cost pass-through estimates. Consider differentiated-products Bertrand competition. Let $\mathcal J_f$ be the set of products owned by firm $f$, let $c$ collect constant marginal costs, and define the ownership matrix $\Omega$, with $\Omega_{ij}=1$ when products $i$ and $j$ are owned by the same firm and zero otherwise. Firm $f$'s variable profit is then:
\begin{equation}
\pi_f(\boldsymbol{p})=(\boldsymbol{p}_f-\boldsymbol{c}_f)'\boldsymbol{q}_f(\boldsymbol{p},X)
\end{equation}

Because $M_{\boldsymbol p,X}^{-1}$ is symmetric, the first-order conditions for all firms can be stacked as follows:
\begin{equation}
\boldsymbol{q}(\boldsymbol{p},X)-\Delta(\boldsymbol{p},X)(\boldsymbol{p}-\boldsymbol{c})=\boldsymbol{0} \qquad \Delta(\boldsymbol{p},X)\equiv\Omega\odot M_{\boldsymbol p,X}^{-1}
\end{equation}

Thus, at any interior optimum:
\begin{equation}
\boldsymbol{p}-\boldsymbol{c}=\Delta(\boldsymbol{p},X)^{-1}\boldsymbol{q}(\boldsymbol{p},X)\quad\Rightarrow\quad \boldsymbol{p}_f-\boldsymbol{c}_f=\{M_{\boldsymbol p,X}^{-1}\}_{|\mathcal{J}_f|\times |\mathcal{J}_f|}^{-1}\boldsymbol{q}_f(\boldsymbol{p},X),\ \forall f
\end{equation}

Note that $\Delta(\boldsymbol{p},X)$ is in most applications a sparse matrix; despite the possibility of higher dimensions for settings with many goods, this property minimises computational concerns.\\

The standard multi-product markup inversion remains available; the only difference is that the relevant demand Jacobian is the corner-adjusted one. Marginal costs can be recovered from observed prices and quantities once $M_{\boldsymbol p,X}^{-1}$ is known, and equilibrium variable profits satisfy:
\begin{equation}
\pi_f=\boldsymbol{q}_f'\{M_{\boldsymbol p,X}^{-1}\}_{|\mathcal{J}_f|\times |\mathcal{J}_f|}^{-1}\boldsymbol{q}_f
\end{equation}

The unconstrained linear case is especially straightforward; it is well-known that optimal multi-product prices are available in closed form. However, with corner solutions, a global closed form is not available because $M_{\boldsymbol p,X}^{-1}$, and so $\Delta(\boldsymbol{p},X)$, changes with prices. Instead, optimal pricing requires solving the following $N$-dimensional system:
\begin{equation}
M^{-1}(X\boldsymbol{\beta}-\boldsymbol{p})+\kappa(\boldsymbol{p},X)
-\Delta(\boldsymbol{p},X)(\boldsymbol{p}-\boldsymbol{c})=\boldsymbol{0}
\end{equation}

What about cost pass-through? It is well-understood that linear demand usually introduces significant restrictions on these. Conditional on $X$ and ownership, unconstrained demand has a constant Jacobian, with equilibrium cost pass-through independent of price and cost levels:
\begin{equation}
\frac{\partial \boldsymbol{p}^*}{\partial \boldsymbol{c}}=\left(M^{-1}+\Delta^I\right)^{-1}\Delta^I
\end{equation}

This inflexibility is undesirable; in the limiting case in which a single firm owns all products, $\Delta^I=M^{-1}$, we get:
\begin{equation}
\frac{\partial \boldsymbol{p}^*}{\partial \boldsymbol{c}}=\frac{1}{2}I
\end{equation}

Corner solutions relax precisely this restriction. Implicitly differentiating the firms' first-order conditions yields:
\begin{equation}
\frac{\partial \boldsymbol{p}}{\partial \boldsymbol{c}}=\left[M_{\boldsymbol p,X}^{-1}+\Delta(\boldsymbol{p},X)+\Xi(\boldsymbol{p},\boldsymbol{c})\right]^{-1}\Delta(\boldsymbol{p},X)
\end{equation}

for $\Xi_{\cdot k}(\boldsymbol{p},\boldsymbol{c})\equiv\frac{\partial\Delta(p,X)}{\partial p_k}(\boldsymbol{p}-\boldsymbol{c})$. Pass-through now depends on the prevailing price vector because a cost shock can move consumers towards or across consumer-level corner boundaries, changing the active demand Jacobian. The aggregate corner correction $\kappa(\boldsymbol{p},X)$ now determines how pass-through. With finitely many consumers, pass-through is locally constant between corner boundaries and changes when a boundary is crossed; with sufficiently rich heterogeneity, these changes are smoothed and pass-through varies continuously with prices.\\

The model consequently retains the tractability of linear demand where it is useful - including a closed-form multi-product pricing benchmark - without imposing constant pass-through globally. Corner solutions are not a panacea however; the resulting pass-through is flexible, but not unrestricted. This is clearest in the single-product case:
\begin{equation}
    \frac{d\boldsymbol{p}}{d\boldsymbol{c}}=\frac{M_{\boldsymbol p,X}^{-1}}{2M_{\boldsymbol p,X}^{-1}-\boldsymbol{q}''(\boldsymbol{p})(\boldsymbol{p}-\boldsymbol{c})}
\end{equation}

Without corners, $\boldsymbol{q}''(\boldsymbol{p})=0$ and the familiar one-half pass-through obtains. With corner solutions, consumers progressively leave the market as price rises, so $\boldsymbol{q}''(\boldsymbol{p})\geq\boldsymbol{0}$: the demand curve locally flattens as additional corners bind. Pass-through can therefore rise above one half, can equal full pass-through, and may exceed one when curvature is sufficiently strong, subject to the second-order condition. Thus, the model does not permit an arbitrary pass-through parameter - for example, the single-product mechanism does not generate pass-through anywhere between zero and one - but it removes the much stronger restriction that pass-through must be constant at one half. As for the the multi-product case, there are no comparable bounds: own- and cross-cost pass-through are jointly determined by $M_{\boldsymbol p,X}^{-1}$, ownership, and by the way corner solutions change the demand Jacobian.\\

These results require no additional estimates. The multi-product pricing system therefore amounts to evaluating these estimated objects and solving the $N$-dimensional first-order conditions. Cost pass-through additionally requires the local derivative of $M_{\boldsymbol p,X}^{-1}$, but this is the same demand curvature identified by our full demand specification with corner corrections. No consumer-level data or separate supply-side demand estimation is required. Given an ownership matrix and Bertrand conduct, marginal costs can themselves be recovered from observed prices and quantities through the markup inversion above; additional supply-side structure is needed only if one wishes to model marginal costs themselves or conduct cost counterfactuals away from the observed equilibrium.

\subsection{Merger analysis}

The model developed in this paper offers an alternative framework for the simulation of horizontal mergers between firms producing differentiated goods. Its main advantages arise from three properties of the demand system: it can be estimated without requiring the researcher to pre-specify a market size; it permits flexible substitution patterns across products; and, through heterogeneous corner solutions, it allows substitution patterns and demand curvature to change endogenously as prices adjust following a merger, including enabling zero demand.\\

Consider a merger between two firms competing in prices. Let $\Omega^0$ be the pre-merger ownership matrix and $\Omega^1$ the post-merger matrix, with the merger changing the ownership indicators for products that were previously controlled by separate firms. Assume initially that product characteristics and marginal costs remain unchanged. The merger affects equilibrium prices because the merged firm internalises substitution between products previously owned by separate firms.\\

From Section \ref{pricingpassthrough}, the pre-merger first-order conditions under differentiated-products Bertrand competition are:
\begin{equation}
q(p^0,X)-\Omega\odot M_{p,X}^{-1}(p^0-c^0)=q(p^0,X)-\Delta(p^0,X;\Omega^0)(p^0-c^0)=0
\end{equation}

Given the estimated demand system and observed pre-merger prices and quantities, marginal costs can be recovered through the standard markup inversion:
\begin{equation}
c^0=p^0-
\Delta(p^0,X;\Omega^0)^{-1}q(p^0,X)
\end{equation}

The post-merger equilibrium can then be calculated by replacing the ownership matrix while holding the estimated demand primitives fixed. Allowing also for possible merger-specific marginal-cost efficiencies, represented by $c^1$, post-merger prices satisfy:
\begin{equation}
M^{-1}(X\beta-p^1)+\kappa(p^1,X)-\Delta(p^1,X;\Omega^1)(p^1-c^1)=0
\end{equation}

When corner solutions are present, the problem requires solving the post-merger first-order conditions while recomputing the aggregate corner correction and the associated demand Jacobian at each candidate price vector. This is economically important. Under the discrete-choice microfoundation developed here, price increases induce different consumers to stop purchasing particular products, changing the set of alternatives between which consumers substitute, and therefore changing the demand Jacobian. Price increases need not be governed by the same local substitution patterns observed before the merger.\\

The second advantage, as mentioned earlier, concerns the treatment of the outside option. BLP estimation requires a measure of the total potential market size to construct the outside-good share. This introduces a modelling choice that can affect the estimated demand response to price changes. The approach proposed here instead lets the demand system be estimated directly using observed aggregate quantities. Substitution towards the outside option is disciplined by the row sums of the estimated demand response matrix, rather than by a separately constructed outside-good market share. The researcher therefore does not need to specify a market size simply to normalise observed quantities into choice probabilities. This property is particularly useful in research, when the researcher observes only a subset of the products available to consumers; and in practice, where merger cases are sometimes decided on the adequacy of a particular market definition.\\

Furthermore, the model allows merger-induced changes in consumer and producer welfare to be assessed using the same estimated demand system. For a merger that changes prices from $p^0$ to $p^1$, holding product characteristics fixed, the change in aggregate consumer surplus is:
\begin{equation}
\Delta W=-\int_0^1q\left(p^0+s(p^1-p^0),X\right)'(p^1-p^0)\ ds
\end{equation}

As established in Section \ref{welfare-analysis}, the discrete-choice microfoundation guarantees that this expression corresponds to a difference in expected surplus and is independent of the integration path. The estimated corner correction enters the welfare calculation directly, allowing the counterfactual to account for consumers who cease purchasing particular products as prices rise.\\

\subsection{Network theory}\label{network-theory}

In this section, let us assume sufficiently rich preference heterogeneity to smooth the consumer-level kinks into continuously differentiable aggregate demand, and consider what aggregate demand looks like for some specific types of unconstrained demand Jacobians. We can lean on the network economics interpretation of $M^{-1}$ as approximately an adjacency matrix, and consider some well-known network graphs. To do so, we will make use of tree-distance matrices, where:
\begin{equation}
    M=\gamma^{d_T(i,j)}
\end{equation}

for $d_T(i,j)$ the number of links on the unique path between nodes $i$ and $j$, the nodes being the goods in the market.\\

The most basic matrix is the complete graph:
\begin{equation}
    d_{complete}(i,j)=1\quad\Rightarrow\quad M_{complete}=
    \begin{pmatrix}
        1&\gamma&\cdots&\cdots&\gamma\\
        \gamma&1&\gamma&\cdots&\gamma\\
        \vdots&\gamma&\ddots&\cdots&\cdots\\
        \vdots&\vdots&\vdots&\ddots&\vdots\\
        \gamma&\gamma&\cdots&\cdots&1
    \end{pmatrix}
\end{equation}

It is equivalent to the assumption that all goods are equally imperfect substitutes. Another common matrix is the Kac–Murdock–Szegö (KMS) matrix (Kac, Murdock, and Szegö, 1953), explored in some depth in Amir, Erickson, and Jin, (2017):
\begin{equation}
    d_{KMS}(i,j)=|i-j|\quad\Rightarrow\quad M_{KMS}=
    \begin{pmatrix}
        1&\gamma&\gamma^2&\cdots&\gamma^{n-1}\\
        \gamma&1&\gamma&\cdots&\gamma^{n-2}\\
        \gamma^2&\gamma&1&\cdots&\gamma^{n-3}\\
        \vdots&\vdots&\vdots&\ddots&\vdots\\
        \gamma^{n-1}&\gamma^{n-2}&\gamma^{n-3}&\cdots&1
    \end{pmatrix}
\end{equation} 

This matrix is equivalent to a line graph, a one dimensional ordering with equal distance $\gamma$ at each segment between alternatives in the one-dimensional ordering. The spokes graph, popularised in Chen and Riordan (2007) and otherwise known as the star graph, can be defined as follows, for good 1 the center node of the graph:
\begin{equation}
    d_{spokes}(i,j)=\begin{cases}
        0&\text{if}\ i=j\\
        1&\text{if}\ i=1,j\neq1\ \text{or}\ j=1,i\neq1\\
        2&\text{if}\ i,j\neq1,i\neq j\\
    \end{cases}
    \quad\Rightarrow\quad M_{spokes}=
    \begin{pmatrix}
        1&\gamma&\gamma&\cdots&\gamma\\
        \gamma&1&\gamma^2&\cdots&\gamma^2\\
        \gamma&\gamma^2&1&\cdots&\gamma^2\\
        \vdots&\vdots&\vdots&\ddots&\vdots\\
        \gamma&\gamma^2&\gamma^2&\cdots&1
    \end{pmatrix}
\end{equation}

This is also known as a core-periphery graph, very popular in network economics. It can be thought of as equivalent to a central good with a unique combination of characteristics, in the product category surrounded by a periphery of goods, each relating to the central good by some unique deviation along one particular characteristic.\\

The spider graph is a related graph that also introduces asymmetry across goods: a central node $c$ with $k$ arms of potentially different lengths $\ell_1,\ldots,\ell_k$. By labeling each non-central node by $(a,r)$, where $a$ identifies its arm and $r$ its distance from the center, we obtain:
\begin{equation}
    \begin{aligned}
    &d_{spider}(c,c)=0\\
    &d_{spider}(c,(a,r))=r\\
    &d_{spider}((a,r),(b,s))=
    \begin{cases}
        |r-s|&\text{if}\ a=b\\
        r+s&\text{if}\ a\neq b
    \end{cases}
    \end{aligned}
    \quad\Rightarrow_{e.g.}\quad
    M_{spider}=
    \begin{pmatrix}
        1&\gamma&\gamma^2&\gamma^3&\gamma^3&\gamma^4\\
        \gamma&1&\gamma&\gamma^2&\gamma^2&\gamma^3\\
        \gamma^2&\gamma&1&\gamma&\gamma&\gamma^2\\
        \gamma^3&\gamma^2&\gamma&1&\gamma^2&\gamma\\
        \gamma^3&\gamma^2&\gamma&\gamma^2&1&\gamma^3\\
        \gamma^4&\gamma^3&\gamma^2&\gamma&\gamma^3&1
    \end{pmatrix}
\end{equation}

The example codified in $M_{spider}$ is one such setting with 6 goods and links $(1,2)$, $(2,3)$, $(3,4)$, $(3,5)$, and $(4,6)$, yielding a central node at 3 and three arms of lengths 2, 2, and 1. The central good (in this case 3) has three direct competitors, while goods 2 and 4 have two each and goods 1, 5, and 6 have only one. Such asymmetries are common in reality, making this a graph of particular interest.\\

Lastly, we consider the bridge graph, which is useful to express product categories with clusters of goods in greater competition within-cluster than between clusters (e.g. bio vs non-bio products, open- vs closed-source software). Let $L$ and $R$ be the sets of nodes on the left and right of the graph respectively. let $c_L\in L$ and $c_R\in R$ be the centers of each side. Defining the depth of each node as $h(i)$, equal to 0 if $i\in \{c_L,c_R\}$, equal to 1 if not, we get the following distance function:
\begin{equation}
\begin{aligned}
    &d_{bridge}(i,j)=\begin{cases}
        0&\text{if}\ i=j\\
        h(i)+h(j)&\text{if}\ i\neq h,i,j\in L\ \text{or}\ i,j\in R\\
        1+h(i)+h(j)&\text{if}\ i\in L, j\in R,\ \text{or vice-versa}
    \end{cases}
\\
    &\quad\Rightarrow_{e.g.}\quad M_{bridge}=
    \begin{pmatrix}
        1&\gamma&\gamma&\gamma&\gamma^2&\gamma^2\\
        \gamma&1&\gamma^2&\gamma^2&\gamma^3&\gamma^3\\
        \gamma&\gamma^2&1&\gamma^2&\gamma^3&\gamma^3\\
        \gamma&\gamma^2&\gamma^2&1&\gamma&\gamma\\
        \gamma^2&\gamma^3&\gamma^3&\gamma&1&\gamma^2\\
        \gamma^2&\gamma^3&\gamma^3&\gamma&\gamma^2&1
    \end{pmatrix}
    \end{aligned}
\end{equation}

The example codified in $M_{bridge}$ is just one such setting with 6 goods, $L=\{1,2,3\}$ and $R=\{4,5,6\}$, with centers at nodes 1 and 4, and links $(1,2)$, $(1,3)$, $(1,4)$, $(4,5)$, and $(4,6)$.\\

Tree-distance matrices can be inverted using the following formula:
\begin{equation}
    M^{-1}=\frac{1}{1-\gamma^2}[\mathrm{diag}(1+(\mathrm{deg}(i)-1)\gamma^2)-\gamma G_t]
\end{equation}

for $G_t$ the implied binary adjacency matrix of each graph, equal to 1 where two nodes share a link and zero otherwise, and $\mathrm{deg}(i)$ the number of links of a given node $i$.\\

For this section, we will work with 6 goods, which implies $6!=720$ possible permutations and $2^6=64$ possible $\mathcal C$. We set $\delta=1$ as the $\lambda_{\pi}$ weights cancel out once we consider every possible permutation. We set $X=I$ (as many goods as product characteristics) and obtain the $\boldsymbol{x}_j'\boldsymbol{\beta}_h=\beta_{hj}$s across all permutations by drawing from a Uniform distribution $U[0.05,0.15]$. We will focus on good 1's aggregate demand function, keeping all other goods' prices fixed at $p_2=\ldots=p_6=0.05$ while varying $p_1$. We will set $\gamma$ at three different points: $0.1,\ 0.15,$ and $0.24$ - higher $\gamma$ raise issues for satisfying the conditions in (\ref{aggregate_conditions}) across some of the networks. Market size is normalised to 1.\\

One important concern is that the presence of zeroes in some of the unconstrained demand Jacobians implied by the networks above. These effectively block certain permutations from existing: if $\{M^{-1}\}_{ij}=0$, then no permutation can have $i$ and $j$ adjacent, as it would imply a positive weight for that permutations' $M_P^{-1}$ and therefore a positive weight for the aggregate $M^{-1}$. To address this, any permutation which implies a forbidden adjacency is truncated, with all products with a higher $|\alpha|$ than that of the good with the lowest $|\alpha|$ of the pair being removed from that permutation. This simulates that permutation's choice set.\footnote{An alternative approach to addressing zeros could be to remove any contradicting permutations outright. Except for the complete network and the KMS line, all orderings have forbidden inside-good adjacencies across the other networks. Using our truncation approach, each network, respectively, has 720, 34, 30, 36, and 36 distinct retained paths.}\\

The results for each network and good 1 are below:

\begin{figure}[H]
    \centering
    \caption{Economic networks and respective aggregate demand graphs for good 1}
    \includegraphics[width=\linewidth]{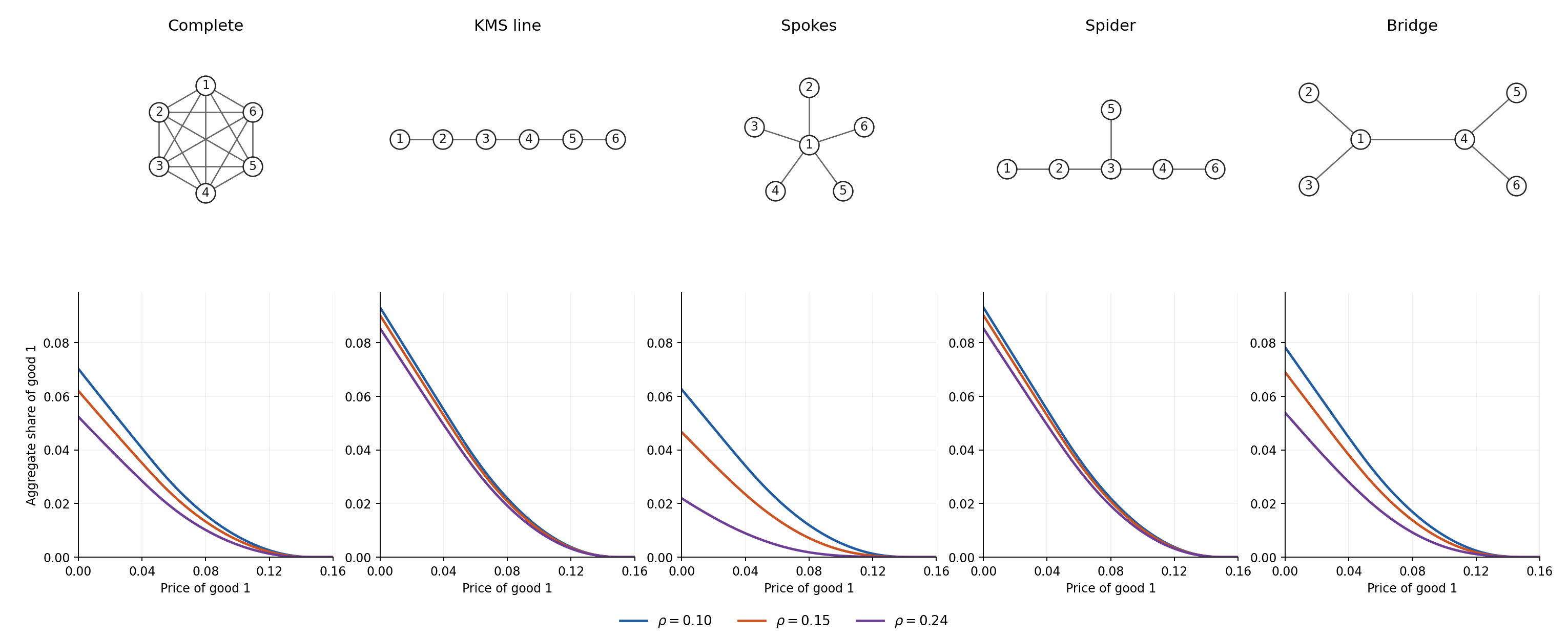}
\end{figure}

Aggregate demand is stated as market shares as we do not specify market size; the use of permutation weights, which sum up to 1, substitutes for consumer weights, which sum up to $H$ and are generally unobserved (as $H$ itself is).\\

The present paper can thus provide theoretical researchers with a way to study variations in market structure with the tools of network theory within a rationalisable model. A strand of the network literature studying welfare-maximising market interventions (Galeotti, Golub, and Goyal, 2020) appears particularly promising for integration. However, the conditions of gross substitution imposed by discrete choice appear to rule out some applications (Galeotti, Golub, Goyal, Talamàs, and Tamuz, 2026).

\subsection{Models of entry and exit}\label{entry-exit}

Market structure - i.e. the number of firms, their sizes, the products they offer in a given market -  and how it ties to market competition is of great interest to IO economists. Many high-quality surveys of this literature exist since at least Bresnahan (1989) (and Berry and Reiss, 2007, more recently), with surveys of stylised facts on entry available since at least Geroski (1995). In the 1990s and early 2000s, several papers on static models of market entry and oligopolistic competition tackled well-known issues with multiplicity of equilibria in simultaneous entry, such as Bresnahan and Reiss (1990, 1991), Berry (1992), Seim (2006), and Ciliberto and Tamer (2009).\\

In the 2010s and 2020s, empirical models of \textit{multi-stage} \textit{portfolio} entry and exit under \textit{multi-product} oligopoly, with the clearest potential for policy and economic research, are the newer development: Eizenberg (2014), Wollmann (2018), Fan and Yang (2025), and Montag (2026) most recently. These papers attempt to endogenise what goods and respective characteristics are introduced by firms onto specific markets, addressing issues related to combinatorial computational complexity. Combinatorial complexity follows from entry-no entry decisions across every possible combination of product characteristics.\\

A fully dynamic game with a fully-specified endogenous product design stage has faced a key bottleneck: a reliance on BLP demand estimation\footnote{Aguirregabiria and Ho (2012) and Rossi (2023) are important exceptions.}. Akchen and Misic (2026) find exact Logit-based product design to be NP-hard, with design instead tackled in Operations Research and Marketing literatures via heuristics and near optimality. In IO, not only does this prevent extensions from multi-stage to dynamic models of firm behaviour, it severely restricts separate identification of concepts of interest, such as endogenous sunk versus fixed fixed costs. The computational complexity of BLP also implies strong restrictions on the number of products one can analyse.\\

The present paper introduces an alternative discrete-choice-based demand system that could alleviate these computational constraints. In particular, the characteristics-based representation of the demand matrix in Lemma 1 allows the number of demand parameters to grow with the number of observed characteristics rather than with the number of products.\\

%Rodrigues (2026) exploits this structure to develop a dynamic model of endogenous product design, entry, and pricing under multi-product oligopoly. The computational gains are substantial whether non-negativity constraints are or not slack. The representation reduces the inversion of the demand matrix to operations on a matrix whose dimension depends only on the number of characteristics, avoiding repeated inversion of large product-level matrices as the choice set changes. Conditional on product characteristics and market structure, Bertrand equilibrium prices and quantities can be obtained through linear algebra rather than repeatedly solving non-linear pricing problems. These gains are particularly consequential in a dynamic setting, where firms must evaluate equilibrium profits and continuation values across numerous candidate designs, entry configurations, and future market states. Moreover, continuous characteristics can be optimised conditional on discrete design choices, considerably reducing the combinatorial burden associated with searching over possible product configurations. While the enumeration of simultaneous entry outcomes and the calculation of continuation values remain computationally demanding, the simplification of the underlying demand and pricing problems makes each branch of the dynamic game substantially less costly to evaluate.

\subsection{Durable goods}\label{durable-goods}

The durable goods literature provides another natural application for aggregate linear demand. Durable goods introduce an inherently dynamic dimension to competition: current sales affect the stock of goods available in future periods, while consumers' expectations about future prices, product quality, and resale values influence current purchasing decisions. The use of linear demand in this literature is not new. Bulow (1982), for example, relies on linear demand to study how the inability of durable-goods monopolists to commit to future production can distort output, investment, and product durability. The challenge has instead been to reconcile the tractability of these theoretical models with the heterogeneous preferences and product differentiation required for empirical analysis.\\

A substantial empirical literature has developed dynamic models of durable-goods markets (see e.g. Esteban and Shum, 2007; Chen, Esteban, and Shum, 2013; Gowrisankaran and Rysman, 2012; Gavazza and Lanteri, 2021). Much like with dynamic entry models, a significant difficulty in extending these models is computational complexity. Studying dynamic firm behaviour requires incorporating changes in installed stocks, demand for product replacements, and implications for prices of new and used goods sold by the firm, as well as those sold by competitors. Interactions must be evaluated repeatedly across different states, candidate action profiles, and counterfactual equilibria. The computational gains from the full model presented in this paper could therefore facilitate richer dynamic models of multi-product competition, endogenous durability, replacement, and secondary-market equilibrium.

\subsection{Retail and large choice sets}\label{retail}

Retail markets provide a particularly demanding setting for structural demand estimation. A supermarket or online retailer may offer hundreds or thousands of differentiated products. At the same time, the most accessible data often consist of market-level data, rather than the choices and complete shopping baskets of individual households over time. The demand model developed here has two especially useful theoretical properties in this setting: estimating the demand for the goods of interest does not require the researcher to construct an outside-option share, and the system can be treated as an incomplete demand system.\\

In conventional discrete-choice applications such as BLP, the researcher typically converts observed sales into shares using a measure of potential market size. For $Market\_Size_t$ one such measure, the outside share is computed as $s_{0t}=1-\sum_{j=1}^{N_t}q_{jt}/Market\_Size_t$. It is often unclear what counts as an appropriate measure. Mismeasurement affects estimated price sensitivity, diversion ratios, markups, pass-through, and welfare counterfactuals (Nevo, 2000; Huang and Rojas, 2014). How this measure is calculated is at the discretion of the researcher, which can raise disputes in real-world merger cases (see e.g. Baker, 2007). In contrast, the demand equation in (\ref{eq:aggregate_structural}) directly relates the quantities of observed inside goods to their prices and characteristics. Estimation does not require assigning a quantity to non-purchase or constructing $s_{0t}$ as an additional dependent variable. The inside-good response matrix already disciplines substitution towards that alternative through (\ref{conditionsagg_2}).\\

The second theoretical property is closely related. Incomplete demand systems can specify demands for a subset of all consumption goods without modelling the entire allocation of the consumer's budget (Epstein, 1982; LaFrance, 1985). In the present model, purchases of the inside products are evaluated against utility from residual income, and I have shown earlier how locally differing marginal utilities before and after purchase allow for income effects without abandoning the linear interior demand structure. The remaining consumption bundle need not be decomposed into a set of priced products and separately estimated demand equations. This is particularly appealing for retail applications: a researcher interested in the price of breakfast cereals should not necessarily have to specify the joint demands for everything else purchased. Current practice sidesteps this entirely, obscuring considerations such as product-specific cross-category effects. Here aggregate linear demand can provide not just potential computational advantages, but enable heterogeneous choice sets across the consumer population, and the building of more robust measures of product substitution and market structure.\\

Furthermore, the model can be extended to the case where retail transactions involve multi-unit, multi-product, cross-category purchases in a single transaction. Let $A$ be an $N\times \text{number of unique baskets}$ product–basket incidence matrix, where each column specifies the quantities of individual products contained in a feasible basket. Consumers now choose among baskets rather than individual products, with basket prices $A' \boldsymbol p$. Applying the same discrete-choice construction at the basket level yields interior basket demand $\boldsymbol{z}(\boldsymbol{p})=M_B^{-1}(X_B\boldsymbol\beta-A' \boldsymbol{p})$, implying aggregate product demand and its Jacobian:

\begin{equation}
\boldsymbol q(\boldsymbol p)=A\boldsymbol z(\boldsymbol p)\qquad\frac{\partial \boldsymbol{q}}{\partial \boldsymbol{p}}=-AM_B^{-1}A'
\end{equation}

Although baskets are substitutes as choice alternatives, individual products may exhibit complementarities through joint purchases. This is one way to maintain the rationalisability conditions imposed by our discrete-choice microfoundation while introducing complementarities of use in cases such as Galeotti, Golub, Goyal, Talamàs, and Tamuz (2026). Corner corrections can likewise be applied at the basket level before aggregation into product demand.\\

Computational tractability would naturally depend on a parsimonious representation of the feasible basket set. On this, Rodrigues (2025) provides a complementary route to aggregate linear demand for these types of decisions via continuous choice. In that setting corner correction is also far more computationally tractable, though it does not identify the distribution of heterogeneous preferences.

\subsection{Economics of platforms}\label{platforms}

The economics of platforms and two-sided markets literature often relies on linear demand systems derived from individual discrete choice to characterise equilibrium prices and study externalities. However, the tractability of these models often depends on full market coverage, where all consumers participate in one of the available platforms. This assumption rules out market expansion: changes in prices reallocate a fixed population across platforms without affecting the total number of participants. Since consumers' decisions to participate in one side of the market affect the attractiveness of platforms to the other side, full coverage is especially problematic for a fuller understanding of the interplay between market size, network effects, and equilibrium prices. Recent contributions continue to face this difficulty, even when analysing alternative demand systems (Peitz and Sato, 2026). \\

The present paper explicitly incorporates an outside option, allowing changes in prices and product characteristics to affect both substitution between platforms and total market participation. When a platform becomes sufficiently unattractive that some consumers exit, the corner correction developed above adjusts the unconstrained demand Jacobian to account for the resulting changes in substitution, including movements towards the outside option itself.

% Dynamic models of entry and exit, flexible cost pass-throughs and markups, applications to large product categories, market definition is no longer a problem, incomplete demand system, extension to endogenous sunk costs, networks, product differentiation/design, platform econ (Armstrong'06), durable goods (Bulow'82).\\

% "Attractive new tool for oligopoly analysis, as well as a representation of monopolistic competition" - i think I can go further, you can model monopoly and perfect competition, and as I show later, vertical and horizontal differentiation (maybe an opportunity to mention product design, as in Rodrigues (2026). Extends to setting without firm or variety symmetry, imperfect substitution, existence and uniqueness of asymmetric price and quantity equilibria follow from from existing results.\\

% "Novel and interesting results about equilibrium price and variety" - more of a potential thing, since we don't have symmetry so it'll all be contextual."\\

% "The model provides an attractive framework to study firm's incentives to offer multiple products and the competition between multi-product firms. (...) market concentration (...) horizontal and vertical mergers (...)" - I would rather target estimation rather than these typical goals of economic theory, thought they are all nested.\\

% \section{Monte Carlo Simulation}

% \section{Conclusion}

\section{References}
\small 

Aguirregabiria, V., \& Ho, C.-Y. (2012). A dynamic oligopoly game of the US airline industry: Estimation and policy experiments. Journal of Econometrics, 168(1), 156–173.\\

Akchen, İ., \& Mišić, V. (2026). Exact logit-based product design. Management Science.\\

Amir, R., Erickson, P., \& Jin, J. (2017). On the microeconomic foundations of linear demand for differentiated products. Journal of Economic Theory, 169, 641–665.\\

Anderson, S., de Palma, A., \& Thisse, J.-F. (1989). Demand for differentiated products, discrete choice models, and the characteristics approach. The Review of Economic Studies, 56(1), 21–35.\\ 

Armstrong, M., \& Vickers, J. (2015). Which demand systems can be generated by discrete choice? Journal of Economic Theory, 158, 293–307.\\

Baker, J. (2007). Market definition: An analytical overview. Antitrust Law Journal, 74(1), 129–173.\\

Berry, S. (1992). Estimation of a model of entry in the airline industry. Econometrica, 60(4), 889–917.\\

Berry, S., Levinsohn, J., \& Pakes, A. (1995). Automobile prices in market equilibrium. Econometrica, 63(4), 841–890.\\

Berry, S., Levinsohn, J., \& Pakes, A. (2004). Differentiated products demand systems from a combination of micro and macro data: The new car market. Journal of Political Economy, 112(1), 68–105.\\

Berry, S., Linton, O., \& Pakes, A. (2004). Limit theorems for estimating the parameters of differentiated product demand systems. The Review of Economic Studies, 71(3), 613–654. \\

Berry, S., \& Reiss, P. (2007). Empirical models of entry and market structure. In M. Armstrong \& R. Porter (Eds.), Handbook of industrial organization (Vol. 3, pp. 1845–1886). Elsevier.\\

Blackwell, D., \& MacQueen, J. (1973). Ferguson distributions via Pólya urn schemes. The Annals of Statistics, 1(2), 353–355.\\

Bos, I., \& Vermeulen, D. (2022). On the microfoundation of linear oligopoly demand. The B.E. Journal of Theoretical Economics, 22(1), 1–15.\\

Bresnahan, T. (1989). Empirical studies of industries with market power. In R. Schmalensee \& R. Willig (Eds.), Handbook of industrial organization (Vol. 2, pp. 1011–1057). Elsevier.\\

Bresnahan, T., \& Reiss, P. (1990). Entry in monopoly market. The Review of Economic Studies, 57(4), 531–553.\\

Bresnahan, T., \& Reiss, P. (1991). Entry and competition in concentrated markets. Journal of Political Economy, 99(5), 977–1009.\\

Bulow, J. (1982). Durable-goods monopolists. Journal of Political Economy, 90(2), 314–332. \\

Chen, J., Esteban, S., \& Shum, M. (2013). When do secondary markets harm firms? American Economic Review, 103(7), 2911–2934.\\

Chen, Y., \& Riordan, M. (2007). Price and variety in the spokes model. The Economic Journal, 117(522), 897–921.\\ 

Choné, P., \& Linnemer, L. (2020). Linear demand systems for differentiated goods: Overview and user’s guide. International Journal of Industrial Organization, 73, Article 102663.\\

Ciliberto, F., \& Tamer, E. (2009). Market structure and multiple equilibria in airline markets. Econometrica, 77(6), 1791–1828.\\

Daly, A., \& Zachary, S. (1978). Improved multiple choice models. In D. Hensher \& M. Dalvi (Eds.), Determinants of travel choice (pp. 321–362). Saxon House.\\

Domencich, T., \& McFadden, D. (1975). Urban travel demand: A behavioral analysis. North-Holland Publishing Company.\\

Eizenberg, A. (2014). Upstream innovation and product variety in the U.S. home PC market. The Review of Economic Studies, 81(3), 1003–1045.\\

Epstein, L. (1982). Integrability of incomplete systems of demand functions. The Review of Economic Studies, 49(3), 411–425.\\

Esteban, S., \& Shum, M. (2007). Durable-goods oligopoly with secondary markets: The case of automobiles. The RAND Journal of Economics, 38(2), 332–354.\\

Fan, Y., \& Yang, C. (2025). Estimating discrete games with many firms and many decisions: An application to merger and product variety. Journal of Political Economy, 133(6), 1886–1931.\\

Fosgerau, M., Melo, E., de Palma, A., \& Shum, M. (2020). Discrete choice and rational inattention: A general equivalence result. International Economic Review, 61(4), 1569–1589.\\

Galeotti, A., Golub, B., \& Goyal, S. (2020). Targeting interventions in networks. Econometrica, 88(6), 2445–2471.\\

Galeotti, A., Golub, B., Goyal, S., Talamàs, E., \& Tamuz, O. (2026). Robust market interventions. Journal of Political Economy.\\

Galichon, A., \& Salanié, B. (2022). Cupid’s invisible hand: Social surplus and identification in matching models. The Review of Economic Studies, 89(5), 2600–2629. \\

Gandhi, A., \& Nevo, A. (2021). Empirical models of demand and supply in differentiated products industries. In K. Ho, A. Hortaçsu, \& A. Lizzeri (Eds.), Handbook of industrial organization (Vol. 4, pp. 63–139). Elsevier.\\

Gavazza, A., \& Lanteri, A. (2021). Credit shocks and equilibrium dynamics in consumer durable goods markets. The Review of Economic Studies, 88(6), 2935–2969.\\

Geroski, P. (1995). What do we know about entry? International Journal of Industrial Organization, 13(4), 421–440.\\

Gowrisankaran, G., \& Rysman, M. (2012). Dynamics of consumer demand for new durable goods. Journal of Political Economy, 120(6), 1173–1219.\\

Grieco, P., Murry, C., \& Yurukoglu, A. (2024). The evolution of market power in the U.S. automobile industry. The Quarterly Journal of Economics, 139(2), 1201–1253.\\

Hofbauer, J., \& Sandholm, W. (2002). On the global convergence of stochastic fictitious play. Econometrica, 70(6), 2265–2294.\\

Hotelling, H. (1929). Stability in competition. The Economic Journal, 39(153), 41–57.\\

Huang, D., \& Rojas, C. (2014). Eliminating the outside good bias in logit models of demand with aggregate data. Review of Marketing Science, 12(1), 1–36.\\ 

Jaffe, S., \& Weyl, E. (2010). Linear demand systems are inconsistent with discrete choice. The B.E. Journal of Theoretical Economics, 10(1).\\

Kac, M., Murdock, W., \& Szegő, G. (1953). On the eigen-values of certain Hermitian forms. Journal of Rational Mechanics and Analysis, 2, 767–800.\\

LaFrance, J. (1985). Linear demand functions in theory and practice. Journal of Economic Theory, 37(1), 147–166.\\

McFadden, D. (1974). Conditional logit analysis of qualitative choice behavior. In P. Zarembka (Ed.), Frontiers in econometrics (pp. 105–142). Academic Press.\\

McFadden, D. (1978). Modelling the choice of residential location. In A. Karlqvist, L. Lundqvist, F. Snickars, \& J. Weibull (Eds.), Spatial interaction theory and planning models (pp. 75–96). North-Holland.\\

McFadden, D., \& Train, K. (2000). Mixed MNL models for discrete response. Journal of Applied Econometrics, 15(5), 447–470.\\

Melitz, M., \& Ottaviano, G. (2008). Market size, trade, and productivity. The Review of Economic Studies, 75(1), 295–316. \\

Montag, F. (2026). Mergers, foreign competition, and jobs: Evidence from the US appliance industry. American Economic Review, 116(6), 2085–2119.\\

Nevo, A. (2000). A practitioner’s guide to estimation of random-coefficients logit models of demand. Journal of Economics \& Management Strategy, 9(4), 513–548.\\

Peitz, M., \& Sato, S. (2026). Asymmetric platform oligopoly. The RAND Journal of Economics, 57(1), 60–77.\\

Pellegrino, B. (2025). Product differentiation and oligopoly: A network approach. American Economic Review, 115(4), 1170–1225.\\

Pinkse, J., Slade, M., \& Brett, C. (2002). Spatial price competition: A semiparametric approach. Econometrica, 70(3), 1111–1153.\\

Rockafellar, R. (1970). Convex analysis. Princeton University Press.\\

Rodrigues, A. (2025). Consumer choice over shopping baskets [Preprint]. arxiv.org/abs/2511.11846\\

Rodrigues, A. (2026). Endogenous product design [Preprint]. arxiv.org/abs/2602.02833\\

Rossi, F. (2023). Mergers with endogenous product choice: The case of the ready-to-eat cereal industry. Quantitative Marketing and Economics, 21(1), 1–64.\\

Salop, S. (1979). Monopolistic competition with outside goods. The Bell Journal of Economics, 10(1), 141–156.\\ 

Seim, K. (2006). An empirical model of firm entry with endogenous product-type choices. The RAND Journal of Economics, 37(3), 619–640.\\

Thomassen, Ø., Smith, H., Seiler, S., \& Schiraldi, P. (2017). Multi-category competition and market power: A model of supermarket pricing. American Economic Review, 107(8), 2308–2351.\\
Tirole, J. (1988). The theory of industrial organization. MIT Press.\\

Wollmann, T. (2018). Trucks without bailouts: Equilibrium product characteristics for commercial vehicles. American Economic Review, 108(6), 1364–1406.\\

\newpage

\appendix

\section{Appendix}
\small\singlespacing

\subsection{Proofs}

\paragraph{Proposition 1.} Fix a consumer $h$ and a matrix of product characteristics $X$. Assume that:
\begin{equation}
\mathbb{E}_{F_h}
[
\max_{i\in\mathcal{N}}
|\varepsilon_i|
\mid X,\boldsymbol{\xi}
]
<\infty,
\end{equation}

that $F_h(\boldsymbol{\varepsilon}\mid X,\boldsymbol{\xi})$ is non-atomic, and that ties occur with probability zero. Define the consumer $h$'s surplus function:
\begin{equation}
\mathcal{V}_h(\boldsymbol{v},X,\boldsymbol{\xi})
=
\int
\max_{i\in\mathcal{N}}
{
v_i+\varepsilon_i
}
\ dF_h(\boldsymbol{\varepsilon}\mid X,\boldsymbol{\xi}),
\end{equation}

where $v_0=0$. Then:
\begin{equation}
\begin{aligned}
\mathcal{V}_h(\boldsymbol{v},X,\boldsymbol{\xi})
&=
\sup_{{A_i}_{i=0}^N}
\{
\sum_{i=0}^N
\int_{A_i}
(v_i+\varepsilon_i)
\ dF_h(\boldsymbol{\varepsilon}\mid X,\boldsymbol{\xi})
\}
\\
&=
\sup_{\boldsymbol{z}:\sum_{i=1}^Nz_i\leq1,,
z_i\geq0\ \forall i=1,\ldots,N}
\
\sup_{{A_i}_{i=0}^N\in\mathcal{P}_h(\boldsymbol{z})}
\{
\sum_{i=0}^N
v_iF_h(A_i\mid X,\boldsymbol{\xi})
+
\sum_{i=0}^N
\int_{A_i}
\varepsilon_i
\ dF_h(\boldsymbol{\varepsilon}\mid X,\boldsymbol{\xi})
\}
\\
&=
\sup_{\boldsymbol{z}:\sum_{i=1}^Nz_i\leq1,,
z_i\geq0\ \forall i=1,\ldots,N}
\{
\boldsymbol{v}'\boldsymbol{z}
+
\sup_{{A_i}_{i=0}^N\in\mathcal{P}_h(\boldsymbol{z})}
[
\sum_{i=0}^N
\int_{A_i}
\varepsilon_i
\ dF_h(\boldsymbol{\varepsilon}\mid X,\boldsymbol{\xi})
]
\}
\\
&=
\sup_{\boldsymbol{z}:\sum_{i=1}^Nz_i\leq1,,
z_i\geq0\ \forall i=1,\ldots,N}
\{
\boldsymbol{v}'\boldsymbol{z}
+
\mathcal{Z}_h(\boldsymbol{z},X,\boldsymbol{\xi})
\}
\\
&=
\max_{\boldsymbol{z}:\sum_{i=1}^Nz_i\leq1,,
z_i\geq0\ \forall i=1,\ldots,N}
\{
\boldsymbol{v}'\boldsymbol{z}
+
\mathcal{Z}_h(\boldsymbol{z},X,\boldsymbol{\xi})
\}.
\end{aligned}
\end{equation}

The first equality follows because the pointwise utility-maximising regions:
\begin{equation}
A_i^*(\boldsymbol{v})
=
\{
\boldsymbol{\varepsilon}:
v_i+\varepsilon_i>v_j+\varepsilon_j,\
\forall j\neq i
\}
\end{equation}

attain the supremum over partitions. Since ties occur with probability zero, these regions form a partition up to a set of measure zero and satisfy:
\begin{equation}
F_h
(
A_i^*(\boldsymbol{v})
\mid X,\boldsymbol{\xi}
)
=
z_{hi}(\boldsymbol{v},X).
\end{equation}

% Therefore, the share vector generated by the individual discrete-choice problem is a maximiser:
% \begin{equation}
% \boldsymbol{z}_h(\boldsymbol{v},X)
% \in
% \arg\max_{\boldsymbol{z}:\sum_{i=1}^Nz_i\leq1,,
% z_i\geq0\ \forall i=1,\ldots,N}
% \{
% \boldsymbol{v}'\boldsymbol{z}
% +
% \mathcal{Z}_h(\boldsymbol{z},X,\boldsymbol{\xi}_h)
% \}.
% \end{equation}

Under the maintained assumptions, the Williams-Daly-Zachary theorem (McFadden, 1978) implies:
\begin{equation}
\nabla_{\boldsymbol{\nu}}
\mathcal{V}_h(\boldsymbol{\nu})
=
\boldsymbol{z}_h(\boldsymbol{\nu}).
\end{equation}

By convex duality, the set of maximisers is the subdifferential of $\mathcal{V}_h$ at $\boldsymbol{\nu}$. Since $\mathcal{V}_h$ is differentiable:
\begin{equation}
\partial_{\boldsymbol{v}}
\mathcal{V}_h(\boldsymbol{\nu})
=\boldsymbol{z}_h(\boldsymbol{\nu})
\end{equation}

Therefore, we can conclude:
\begin{equation}
\boldsymbol{z}_h(\boldsymbol{\nu})=
\arg\max_{\boldsymbol{z}:\sum_{i=1}^Nz_i\leq1,,
z_i\geq0\ \forall i=1,\ldots,N}
\{
\boldsymbol{\nu}'\boldsymbol{z}
+
\mathcal{Z}_h(\boldsymbol{z},\boldsymbol{\nu})
\}.
\end{equation}\qedwhite

\paragraph{Proposition 2.} For an individual consumer, how a price increase affects their choice of alternative can be written, in matrix notation:
\begin{equation}
    D_pd_{h}^M=D_p^+d_{h}^H-(D_yd_{h}^M)d_{h}^{M'}
\end{equation}

for $d_{h}$ the $N\times 1$ vector of individual choice indicators, with at most one non-zero element on the $i^*_r$-th row; $D_p^+$ the right-hand compensated price derivative; and superscripts referring to Marshallian and Hicksian terms. Since $d_{h}$ is discrete, these are not ordinary pointwise derivatives; they are zero away from indifference boundaries and undefined at the boundary. The equation above should therefore be understood in terms of boundary changes evaluated at the pre-change choice. By taking expectations on these boundary contributions over $F(\varepsilon)$, we can however obtain consumer $h$'s Slutsky matrix:
\begin{equation}
    D_p\boldsymbol{z}_h^M=\underbrace{D_p\boldsymbol{z}_h^H}_{\text{substitution effect}}-\underbrace{\mathbb{E}[(D_yd_{h}^M)d_{h}^{M'}]}_{\text{income effect}}
\end{equation}

Notice that $\boldsymbol{z}_h$ cannot be obtained separately in the income effect term, which explains its absence in (\ref{slutskyeq}). Let us characterise $\mathbb{E}[(D_yd_{h}^M)d_{h}^{M'}]$. From (\ref{seglevelchoiceprob}), we obtain:
\begin{equation}
    \mathbb E\left[\frac{\partial d_{hi}^M}{\partial y_{h}}\right]
    =\frac{\partial z_{hi}}{\partial y_{h}}=\frac{\partial z_{hi}}{\partial y_{h}}\frac{\partial y_{h}}{\partial y_{h}}=\frac{\partial z_{hi}}{\partial y_{h}}=(\lambda_{h}^1-\lambda_{h}^0)(M_h^{-1}\boldsymbol 1)_i
\end{equation}
with
\begin{equation}\label{insideoutside}
    \mathbb E\left[\frac{\partial d_{hi}^M}{\partial y_{h}}d_{hrj}^M\right]
    =
    \begin{cases}
    (\lambda_{h}^1-\lambda_{h}^0)
    (M_h^{-1}\boldsymbol 1)_i
        & \text{if }i=j\\
    0 & \text{otherwise}
    \end{cases}
\end{equation}

In matrix form:
\begin{equation}
    \mathbb E\left[
        (D_y d_{h}^M)d_{h}^{M'}\right]
    =
    (\lambda_h^1-\lambda_h^0)
    \cdot\operatorname{diag}(M_h^{-1}\boldsymbol 1)
\end{equation}

Substituting into the Slutsky equation gets us the compensated response to a marginal increase $D_p \boldsymbol{z}_h^H$, our substitution effect:
\begin{equation}
    \begin{aligned}
    &\ D_p\boldsymbol{z}_h^M=D_p^+\boldsymbol{z}_h^H-\mathbb{E}[(D_yd_{h}^M)d_{h}^{M'}]\\
    \Leftrightarrow&\ D_p \boldsymbol{z}_h^M=D_p^+ \boldsymbol{z}_h^H-(\lambda_h^1-\lambda_h^0)\cdot\operatorname{diag}(M_h^{-1}\boldsymbol 1)\\
    \Leftrightarrow&\ -\lambda_h^1M_h^{-1}=D_p^+ \boldsymbol{z}_h^H-(\lambda_h^1-\lambda_h^0)\cdot\operatorname{diag}(M_h^{-1}\boldsymbol 1)\\
    \Leftrightarrow&\ D_p^+ \boldsymbol{z}_h^H=-\lambda_h^1M_h^{-1}+(\lambda_h^1-\lambda_h^0)\cdot\operatorname{diag}(M_h^{-1}\boldsymbol 1)
    \end{aligned}
\end{equation}

Aggregating across consumers yields (\ref{slutskyeq}). This completes the proof.\footnote{To see the converse result for a price decrease, note that, at the pre-change choice, $d_{hri}^M=0$, meaning $\mathbb{E}[(D_yd_{hr}^M)d_{hr}^{M'}]=0$, and thus $D_p^- z_h^H=-\lambda_h^1M_h^{-1}=D_pz_h^M$ - no income effect in that instance.}\qedwhite\\

\paragraph{Proposition 3.} Since $0\preceq M_{\mathcal C_h}^{-1}\preceq M_h^{-1}$, the consumer-level price responses, including across changes in the binding constraints, are bounded in norm:
\begin{equation}
    \sum_h\mu_h||M_h^{-1}||\leq\sum_h\mu_h\mathrm{tr}(M_h^{-1})=\mathrm{tr}(M^{-1})<\infty
\end{equation}

The last inequality follows from $M^{-1}$ being a well-defined finite matrix.\\

Now consider any sequence of prices $\boldsymbol p_n\to\boldsymbol p$. As per the proposition statement, at any given prices, almost every consumer is sufficiently far from a change in its binding constraints. In other words, for almost every consumer $h$, sufficiently small price changes do not alter $\mathcal C_h$, such that:
\begin{equation}
M_{\mathcal C_h(\boldsymbol p_n,X)}^{-1}
\longrightarrow M_{\mathcal C_h(\boldsymbol p,X)}^{-1}
\end{equation}

Since these matrices are bounded away from infinity. By the dominated convergence theorem, where we may interchange a limit and an integral provided the quantities inside the integral are bounded by an integrable function, we have the necessary conditions to differentiate under aggregation:
\begin{equation}
    \lim_{n\to\infty}D_p\boldsymbol q(\boldsymbol p_n,X)=-\sum_h\mu_h\lim_{n\to\infty}M_{\mathcal C_h(\boldsymbol p_n,X)}^{-1}=D_p\boldsymbol q(\boldsymbol p,X)
\end{equation}

Thus, the aggregate demand Jacobian is continuous.\qedwhite\\

\paragraph{Proposition 4.} To see this, define said matrix's full $(N+1)\times(N+1)$ response matrix:
\begin{equation}
\begin{aligned}
L^{-1}&=
\begin{pmatrix}
\boldsymbol1'M^{-1}\boldsymbol1&-(M^{-1}\boldsymbol1)'\\
-M^{-1}\boldsymbol1&M^{-1}
\end{pmatrix}
=\sum_{0\leq i<j\leq N}
w_{ij}
(\boldsymbol e_i-\boldsymbol e_j)
(\boldsymbol e_i-\boldsymbol e_j)',
\qquad
w_{ij}\equiv-\{L^{-1}\}_{ij}>0.
\end{aligned}
\end{equation}

Now consider the set of all possible one-dimensional orderings $\mathcal P$ of the $N+1$ alternatives, and let $n_{ij}$ be the number of such orderings in which $i$ and $j$ are adjacent.\footnote{Note that if any element in $M^{-1}$ is zero, then not all possible one-dimensional orderings should be used: the respective pair of goods pertaining to that element cannot be adjacent in any permutation we simulate.} Choose any positive weights $\lambda_P$ satisfying $\sum_P\lambda_P=1$. For every pair $i,j$ that is adjacent in ordering $\mathcal P$, choose their distance on that consumer's ordering to satisfy:
\begin{equation}
d_{P,ij}
=
\frac{\lambda_P n_{ij}}{w_{ij}}
\qquad\Leftrightarrow\qquad
\frac{1}{d_{P,ij}}
=
\frac{w_{ij}}{\lambda_P n_{ij}}
\end{equation}

The distance between adjacent alternatives as defined by a given consumer will be increasing in the ordering weight, the number of alternatives, and decreasing in their desired substitution weight. The response matrix generated by ordering $\mathcal P$ is therefore:
\begin{equation}
L^{-1}_P
=
\sum_{{i,j}\text{ adjacent in }P}
\frac{w_{ij}}{\lambda_P n_{ij}}
(\boldsymbol e_i-\boldsymbol e_j)
(\boldsymbol e_i-\boldsymbol e_j)'.
\end{equation}

Aggregating across consumers gives:
\begin{equation}
\begin{aligned}
\sum_P\lambda_P L_P^{-1}
&=
\sum_P
\sum_{{i,j}\text{ adjacent in }P}
\frac{w_{ij}}{n_{ij}}
(\boldsymbol e_i-\boldsymbol e_j)
(\boldsymbol e_i-\boldsymbol e_j)'
\\
&=
\sum_{0\leq i<j\leq N}
w_{ij}
(\boldsymbol e_i-\boldsymbol e_j)
(\boldsymbol e_i-\boldsymbol e_j)'
\\
&=
L^{-1}
\end{aligned}
\end{equation}

The second equality follows because each pair $i,j$ is adjacent in exactly $n_{ij}$ of the orderings. Taking the inside-good block, we recover $M^{-1}=\sum_P\lambda_PM_P^{-1}$. \\

\end{document}